# The Diffraction of Microparticles on Single-layer and Multi-layer Statistically Uneven Surfaces

**Mikhail Batanov-Gaukhman**[1]

Ph.D., Associate Professor, Institute No. 2 "Aircraft, rocket engines and power plants",
Federal State Budgetary Educational Institution of Higher Education "Moscow Aviation Institute
(National Research University)", Volokolamsk highway 4, Moscow, Russian Federation

**Abstract:** In this article: *a*) a method is developed for calculating volumetric diagrams of elastic scattering of microparticles (in particular, electrons and photons) on single-layer and multi-layer statistically uneven surfaces; *b*) the diffraction of elementary particles on crystals is explained without involving de Broglie's idea of the wave properties of matter; *c*) the probability density functions of the derivative of various stationary random processes are obtained; *d*) volumetric diagrams of the scattering of particles and photons on homogeneous and isotropic uneven surfaces with Gaussian, uniform, Laplace, sinusoidal, and other distributions of unevenness are obtained.

**Keywords:** electron diffraction on a crystal, particle scattering diagram, wave scattering diagram, volumetric scattering diagram, statistically uneven surface, derivative of a stationary random process, Kirchhoff approximation

---

[1] alsignat@yandex.ru

**1 Background and introduction**

In 1924, Louis de Broglie suggested that a uniformly and rectilinearly moving particle with mass $m$ and velocity $v$ can be associated with a plane wave

$$\psi = exp\{i(Et - \mathbf{pr})/h\}, \quad (1.1)$$

where $E$ is the kinetic energy of the particle; $\mathbf{p} = m\mathbf{v}$ is its momentum; $h$ is Planck's constant.

The length of such a monochromatic wave is determined by the de Broglie formula

$$\lambda_b = h/mv. \quad (1.2)$$

This idea served as the basis for the development of wave-particle duality and, in particular, made it possible to explain a number of experiments on the diffraction of electrons, neutrons, and atoms by crystals and thin films [1, 2]. Since then, it has been assumed that the diffraction maxima in the Dewisson and Germer experiment appear in directions that meet the Wolfe - Bragg condition $2d \sin \theta_s = n\lambda_{eb}$, or taking into account the refraction of de Broglie's "electronic waves" in a crystal [1]:

$$2d\left(n_e^2 - \cos^2 \theta_s \right)^{\frac{1}{2}} = n\lambda_{eb}, \quad (1.3)$$

where $d$ is the interplanar distance of the crystal lattice, $\theta_s$ is the Bragg's glancing angle (Figure 1), $n = 1, 2, 3 ...$ is the order of interference (or reflection), $\lambda_{eb}$ is the de Broglie electron wavelength, $n_e$ is the refractive index of the de Broglie electron wave.

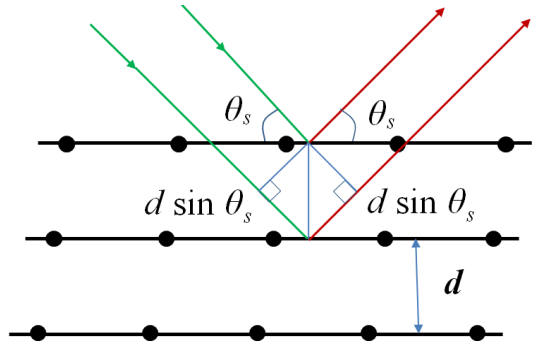

**Fig. 1** Wulff - Bragg's condition for diffraction of microparticles (in particular, electrons or photons) on the surface of a crystal. ⟶ is the direction of motion of the falling microparticles; ⟶ is the direction of movement of the reflected microparticles

However, over the past 95 years, de Broglie waves have not been detected experimentally. They remained an auxiliary mental construction, which allows one to describe the phenomenon mathematically, without revealing the essence of the events occurring in this case.

This article shows that the diffraction of microparticles on a crystal can be described without involving de Broglie's idea of the wave properties of matter.

Based on the laws of reflection in geometric optics and probability theory, at the end of this article, the formula (3.9) [or (3.10)] is obtained for calculating the diagram of elastic scattering of microparticles (DESM) on a multilayer crystal surface. The results of calculations using this formula are consistent with experimentally obtained electron diffraction patterns (EDP) (Figure 1a).

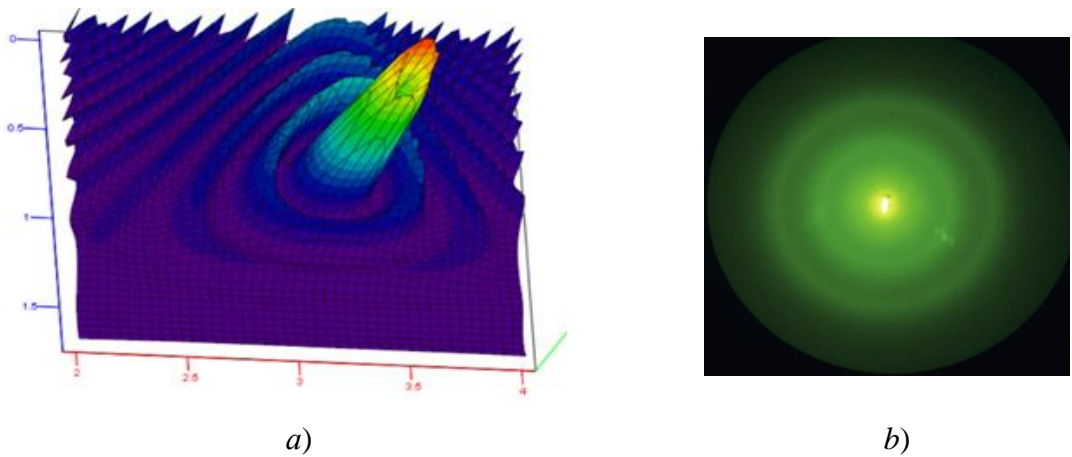

*a)*          *b)*

Fig. 1a *a*) The volumetric diagram of elastic scattering of microparticles on a multilayer crystal surface, obtained as a result of calculations by the formula (3.9); *b*) Experimentally obtained electron diffraction pattern (photo from https://www.sciencephoto.com/media/3883/view)

In addition, this article develops a method for calculating volumetric diagrams of elastic scattering of microparticles (DESM) on uneven surfaces with various statistics of the heights of the irregularities.

By "microparticles" in this paper we mean any particles (fermions and bosons) whose sizes (or wavelength) are much smaller than the characteristic sizes of the irregularities of the reflecting surface (Kirchhoff approximation), and whose reflection occurs according to the laws of geometric optics.

For example, an electron can be called a "microparticle" with an effective size of about $10^{-13}$cm, which is reflected from the surface of a crystal with characteristic sizes of irregularities greater than $10^{-11}$cm. Also, a football can be considered a "microparticle" with a diameter of about 22.3 cm, reflected from an uneven solid surface, the average radius of curvature of which is more than 20 m. "Microparticles" also include photons and phonons with a wavelength $\lambda$ two orders of magnitude smaller than the autocorrelation radius of the heights of the reflecting surface irregularities (Appendix 1).

Extensive literature is devoted to the scattering of particles and waves on the uneven (rough) boundary of two media, for example, [3 – 27]. However, the formulas for calculating volumetric diagrams of scattering of particles or waves on surfaces with different statistics of roughness (irregularities) heights in the case of the Kirchhoff approximation are practically absent in the literature.

Extensive literature is devoted to the scattering of particles and waves on the uneven (rough) boundary of two media, for example, [3 – 27]. However, the formulas for calculating volumetric diagrams of scattering of particles or waves on surfaces with different statistics of roughness (irregularities) -- in the case of the Kirchhoff approximation -- are practically absent in the literature.

This article presents for the first time volumetric diagram of elastic scattering of microparticles on homogeneous and isotropic uneven surfaces with Gaussian, uniform, Laplace, sinusoidal, and other distributions of height of irregularities. Data from the scattering diagram refer to any of the above microparticles (fermions and bosons).

## 2 Method

The purpose of this section of the article is to develop a method for calculating volumetric diagrams of elastic scattering of microparticles (DESM) (in particular, electrons, photons or phonons) on statistically uneven surfaces under the conditions of the Kirchhoff approximation (i.e., when the averaged radius of curvature or the radius of autocorrelation of irregularities reflecting surface is much larger than the size or wavelength of the microparticles).

**2.1 Reflection of elastic microparticles from an uneven surface**

Consider the incidence of microparticles on the surface of a solid (or liquid) body (Figure 2) at the angles $\vartheta$ and $\gamma$ (Figure 3), and their reflection from this surface at the angles $v$ and $\omega$.

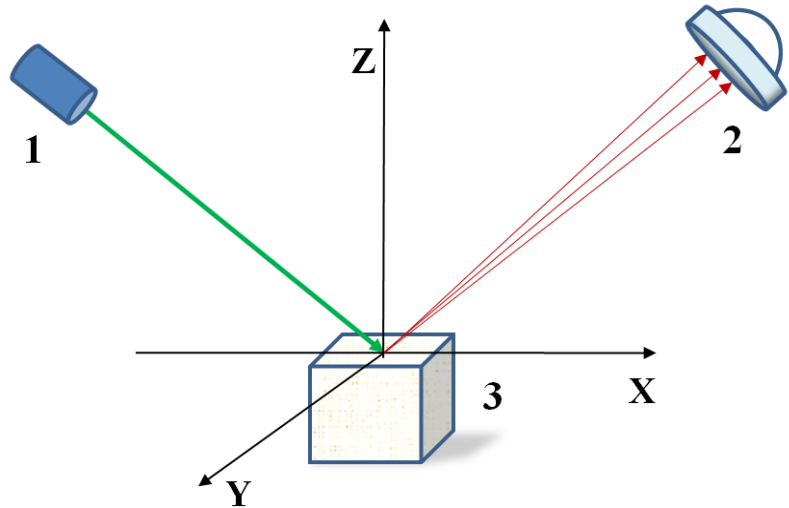

**Fig. 2** Scattering of microparticles (in particular, electrons or photons, that is, a ray of light) on a reflective surface, where: 1 is a microparticle generator; 2 is microparticle detector; 3 is solid or liquid body (in particular, a metal crystal or a volume of water)

Let's imagine the upper layer of a body as a two-dimensional statistically uneven surface $\xi(x,y)$, which repeats the structure of its atomic lattice (Figure 3), or the excitement of a liquid.

It is known that elastic particles (or waves) moving at the speed of v are reflected from the smooth surface of a solid (or liquid) body according to the laws of geometric optics (specular reflection): 1) the incident particle (or light ray), reflected particle (or light ray) and the perpendicular (normal) to the two media border in the point of particle (or light ray) incidence lie in one plane; 2) the angle of incidence $Q_1$ is equal to angle of reflection $Q_2$. This phenomenon is called "specular reflection" or "elastic scattering" of microparticles.

Under the condition of elastic scattering, the fact that the two-dimensional probability density function (TPDF) $\rho(v,\omega)$ of the microparticle is reflected from the uneven reflecting surface at angles $v$, $\omega$ and corresponds to the fact that the TPDF $\rho(\chi,\varphi)$ of the unit vector normal to the surface **n**, at the point where the microparticle falls, will be directed at angles $\chi$, $\varphi$ (Figures 3,4,5).

Therefore, from the TPDF $\rho(\chi,\varphi)$ it is possible to obtain the TPDF $\rho(v,\omega/\vartheta,\gamma)$ using the transformation of variables:

$$\rho(\chi,\varphi) = \rho\{\chi=f_1(v,\omega/\vartheta,\gamma); \ \varphi=f_2(v,\omega/\vartheta,\gamma)\} |G_{v\omega}| = \rho(v,\omega/\vartheta,\gamma)|G_{v\omega}|, \qquad (2.1)$$

where

$$\chi = f_1(v,\omega/\vartheta,\gamma) \qquad (2.2)$$

is the functional relationship between the angle $\chi$ and the angles $v$, $\omega$, for given angles $\vartheta$, $\gamma$;

$$\varphi = f_2(v,\omega/\vartheta,\gamma) \qquad (2.3)$$

is the functional relationship between the angle $\varphi$ and the angles $v$, $\omega$, for given angles $\vartheta$, $\gamma$;

$\rho\{v,\omega/\vartheta,\gamma\}|G_{v\omega}|$ is the TPDF of the microparticle which is reflected from an uneven surface in the direction given by the angles $v$, $\omega$, if angles $\vartheta$, $\gamma$ are known;

$|G_{v\omega}|$ is the Jacobian of the transformation of the variables $\chi$, $\varphi$ into the variables $v$, $\omega$.

In this case, the probability that a particle whose initial direction of motion is given by the angles $\vartheta$ and $\gamma$ will be reflected from the surface in a direction limited by the angular ranges $dv$ and $d\omega$ is

$$P(v,\omega) = \rho(v,\omega/\vartheta,\gamma)|G_{v\omega}|dvd\omega.$$

This formula essentially shows what portion of the total number of microparticles (or the total wave energy) that fall on the reflective surface is scattered in the direction given by the angles $v$ and $\omega$ within the element of the solid angle $d\Omega=dvd\omega$.

If the generator and the detector of microparticles are located at a large distance from the considered part of the reflecting surface (Figure 2), then the TPDF $\rho(v,\omega/\vartheta,\gamma)|G_{v\omega}|$ (2.1) determines the volumetric diagram of elastic scattering of these particles on this surface

$$D(v,\omega/\vartheta,\gamma) = \rho(v,\omega/\vartheta,\gamma)|G_{v\omega}|. \qquad (2.4)$$

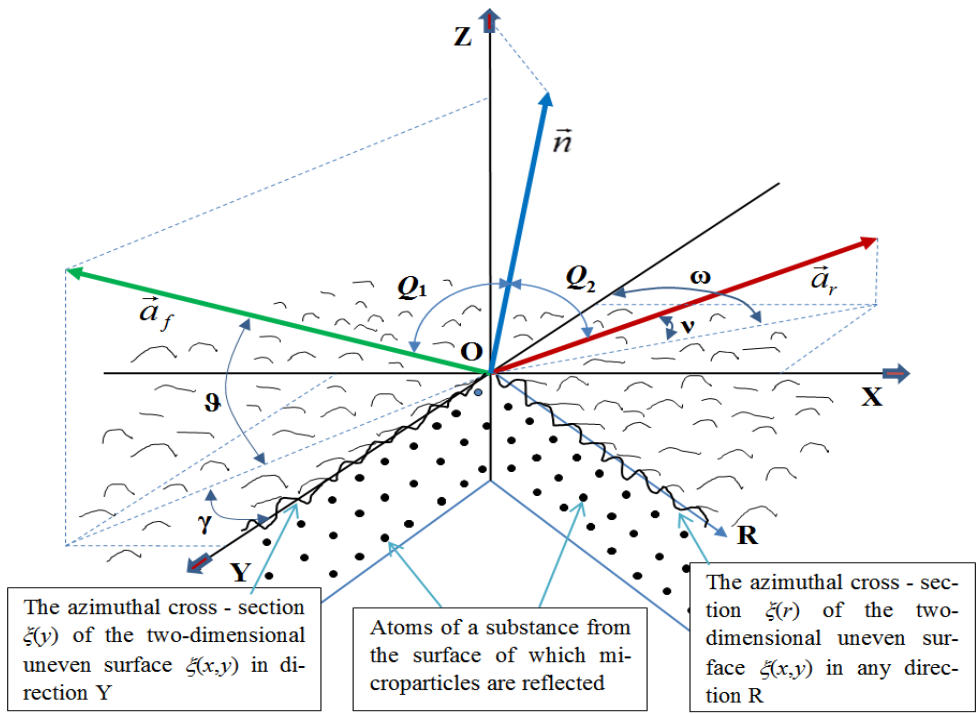

**Fig. 3** Area of uneven surface, reflecting microparticles, where:
$\vartheta, \gamma$ are the angles defining the direction of microparticle incidence on a reflecting surface;
$v, \omega$ are the angles defining the direction of reflection of the microparticle from this surface;
**a**$_f$ is a unit vector indicating the direction to the microparticle generator;
**n** is the unit normal vector to the surface at the point where the microparticle incidence;
**a**$_r$ is a unit vector indicating the direction of motion of the microparticle after an elastic collision with a reflective surface

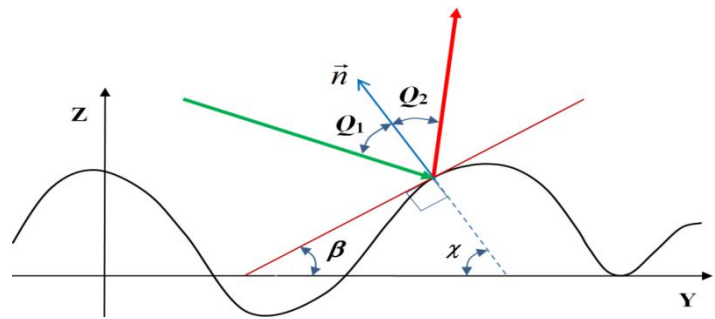

**Fig. 4** Elastic (specular) reflection of a microparticle from a portion of an uneven surface according to the laws of geometric optics: 1) elastic reflection of a particle (or light ray) occurs in the plane of its incidence; 2) the angle of reflection of the particle (or ray of light) $Q_2$ is equal to the angle of its incidence $Q_1$ (i.e., the condition $Q_2 = Q_1$ is satisfied)

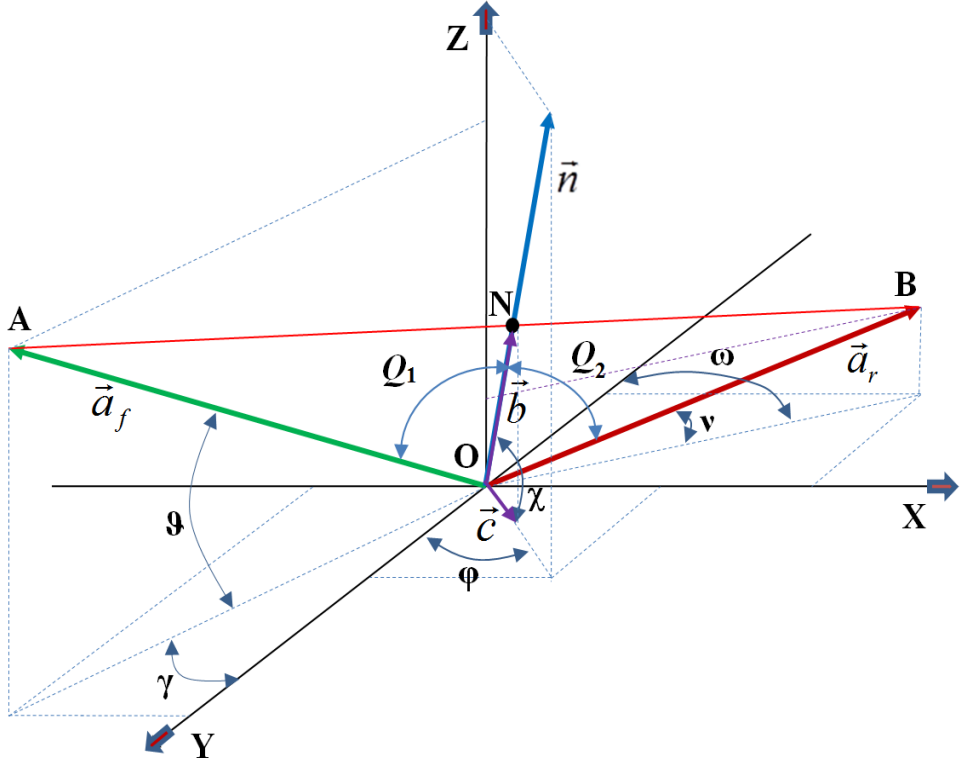

**Fig. 5** Illustration for determining the functional relationship between angles $\chi$, $\varphi$ and angles $\nu$, $\omega$, if angles $\vartheta$, $\gamma$ are known: where $Q_2 = Q_1$ and vectors $\vec{a}_f, \vec{a}_r, \vec{n}$ lie in the same plane AOB

In the case where the microparticle detector is located at a short distance from the considered part of the reflecting surface, then in order to find the volumetric DESM $D(\nu,\omega/\vartheta,\gamma)$, the right part of Expression (2.4) should be integrated over all angles $\nu$ and $\omega$, along which reflected microparticles can enter the detector aperture (or at one point on the plate of the electron diffraction pattern or radiographs).

$$P(\nu,\omega/\vartheta,\gamma) = \int_{\nu_1}^{\nu_2}\int_{\omega_1}^{\omega_2} \rho(\nu,\omega/\vartheta,\gamma)|G_{\nu\omega}|d\nu\,d\omega. \qquad (2.5)$$

This case is not considered in this article. That is, in the future we will assume that the generator and the detector of microparticle are so far from the reflecting surface area that it is permissible to use the simplified formula (2.4).

## 2.2 Functional relationship between the angles $\chi, \varphi$ and $\nu, \omega/\vartheta, \gamma$

Let's find the functional relationships (2.2) and (2.3). Figure 5 shows the unit vectors $\vec{a}_f, \vec{a}_r, \vec{n}$, whose tails coincides with the origin of the local reference system XYZ (located at the point of collision of the microparticle with the surface), and their heads are given by the following coordinates:

$$\vec{a}_f = \{a_{fx}, a_{fy}, a_{fz}\} = \{\cos\vartheta \sin\gamma, \cos\vartheta \cos\gamma, \sin\vartheta\} \tag{2.6}$$

–- a unit vector, indicating the direction on the microparticle generator (Figure 2 and 3);

$$\vec{a}_r = \{a_{rx}, a_{ry}, a_{rz}\} = \{\cos\nu \sin\omega, \cos\nu \cos\omega, \sin\nu\} \tag{2.7}$$

– a unit vector, indicating the direction of movement of a microparticles after an elastic collision with a reflecting surface.

$$\vec{n} = \{n_x, n_y, n_z\} = \{\cos\chi \sin\varphi, \cos\chi \cos\varphi, \sin\chi\}$$

– a unit normal vector to the surface at the point of incidence of the microparticle;

Figure 5 shows that when the laws of geometric optics are satisfied (i.e., when $Q_2 = Q_1$), the normal vector **n** determines the direction of the bisector of the isosceles triangle AOB whose sides are the unit vectors $\mathbf{a}_f$ and $\mathbf{a}_r$,

Obviously by setting the coordinates of the point $N$ that divides the segment AB in half, we get the coordinates of the head of the vector **b**, the direction of which coincides with that of the normal vector **n**. Using the coordinates of the head of the vector $\mathbf{a}_r$ (2.6) and the head of the vector $\mathbf{a}_r$ (2.7), and based on the methods of analytical geometry [29 - 31], we obtain

$$\vec{b} = \{b_x, b_y, b_z\} = \left\{\frac{\cos\nu \sin\omega + \cos\vartheta \sin\gamma}{2}, \frac{\cos\nu \cos\omega + \cos\vartheta \cos\gamma}{2}, \frac{\sin\nu + \sin\vartheta}{2}\right\}.$$

From the scalar product of the vectors **b** and $\mathbf{k} = \{0,0,1\}$ (where **k** indicates the direction of the OZ axis, see Figure 5) $(\vec{b} \cdot \vec{k}) = |\vec{b}| \cdot |\vec{k}| \cos(\pi/2 - \chi) = |\vec{b}| \cdot |\vec{k}| \sin\chi$, we define the angle $\chi$

$$\chi = \arcsin\left(\frac{(\vec{b} \cdot \vec{k})}{|\vec{b}| \cdot |\vec{k}|}\right) = \arcsin\left(\frac{\sin\nu + \sin\vartheta}{\sqrt{(\cos\nu \sin\omega + \cos\vartheta \sin\gamma)^2 + (\cos\nu \cos\omega + \cos\vartheta \cos\gamma)^2 + (\sin\nu + \sin\vartheta)^2}}\right), \tag{2.8}$$

which is the desired functional relationships (2.2).

Figure 5 shows that $\varphi$ is the angle between vectors **j** = {0,1,0} and **c**, where **j** defines the direction of the axis *OY*, and vector **c** is the projection of vector **b** onto the *XOY* plane.

$$\vec{c} = \{c_x, c_y, 0\} = \left\{ \frac{\cos\nu \sin\omega + \cos\vartheta \sin\gamma}{2}, \frac{\cos\nu \cos\omega + \cos\vartheta \cos\gamma}{2}, 0 \right\}. \qquad (2.9)$$

From the scalar product of the vectors $(\vec{c} \cdot \vec{j}) = |\vec{c}| \cdot |\vec{j}| \cos\varphi$, we define the angle $\varphi$

$$\varphi = \arccos\left( \frac{(\vec{c} \cdot \vec{j})}{|\vec{c}| \cdot |\vec{j}|} \right) = \arccos\left( \frac{\cos\nu \cos\omega + \cos\vartheta \cos\gamma}{\sqrt{(\cos\nu \sin\omega + \cos\vartheta \sin\gamma)^2 + (\cos\nu \cos\omega + \cos\vartheta \cos\gamma)^2}} \right), \qquad (2.10)$$

which is the second desired functional relationship (2.3).

## 2.3 Jacobian of the transformation of the variables $\chi, \varphi$ into the variables $\nu, \omega$ for given $\vartheta, \gamma$

Let's introduce the notations

$a = \cos\nu \cos\omega + \cos\vartheta \cos\gamma; \quad b = \cos\nu \sin\omega + \cos\vartheta \sin\gamma; \quad d = \sin\nu + \sin\vartheta; \quad a_\nu' = -\sin\nu \cos\omega;$

$b_\nu' = -\sin\nu \sin\omega; \quad c_\nu' = \cos\nu; \quad a_\omega' = -\cos\nu \sin\omega; \quad b_\omega' = \cos\nu \cos\omega.$ (2.11)

In this case, expressions (2.8) and (2.10), taking into account the polysemy of inverse trigonometric functions, take the form

$$\chi = m\pi + (-1)^m \arcsin\left( \frac{d}{\sqrt{a^2 + b^2 + d^2}} \right), \qquad (2.12)$$

$$\varphi = 2\pi m \pm \arccos\left( \frac{a}{\sqrt{a^2 + b^2}} \right), \qquad (2.13)$$

where $m = 0, 1, 2, 3, \ldots$

From Figures 2, 3, 5 it can be seen that the angles $\chi, \varphi$ can take the values $\chi \in [0, \pi/2]$, $\varphi \in [0, \pi]$. Let's also take into account that the principal branches of inverse trigonometric functions are enclosed within: $\arcsin(x) \in [-\pi/2, \pi/2]$, $\arccos(x) \in [0, \pi]$. Therefore, in expressions (2.12) and (2.13) we assume $m = 0$ and choose (+), and as a result we obtain unambiguous functional relationships

$$\chi = f_1(v, \omega/\vartheta, \gamma) = \arcsin\left(\frac{d}{\sqrt{a^2 + b^2 + d^2}}\right), \quad (2.14)$$

$$\varphi = f_2(v, \omega/\vartheta, \gamma) = \arccos\left(\frac{a}{\sqrt{a^2 + b^2}}\right). \quad (2.15)$$

Let's find the Jacobian of the transformation $|G_{v\omega}|$ variables $\chi$, $\varphi$ into variables $v$, $\omega$. To do this, we calculate the determinant of the matrix [32, 33]

$$|G_{v\omega}| = \begin{vmatrix} \frac{\partial f_1}{\partial v} & \frac{\partial f_1}{\partial \omega} \\ \frac{\partial f_2}{\partial v} & \frac{\partial f_2}{\partial \omega} \end{vmatrix} = \frac{\partial f_1}{\partial v}\frac{\partial f_2}{\partial \omega} - \frac{\partial f_1}{\partial \omega}\frac{\partial f_2}{\partial v}. \quad (2.16)$$

Substituting functions (2.14) and (2.15) into the determinant (2.16), and taking into account the notation (2.11), we obtain the desired Jacobian of the transformation

$$|G_{v\omega}| = \left|\frac{d(a'_v b'_\omega - a'_\omega b'_v) + c'_v(ba'_\omega - ab'_\omega)}{\sqrt{a^2 + b^2}(a^2 + b^2 + d^2)}\right|. \quad (2.17)$$

This result was obtained by the author together with Dr. S.V. Kostin.

**2.4 Definition of TPDF $\rho(\chi,\varphi)$**

Let's obtain the TPDF $\rho(\chi,\varphi)$ (2.1). To do this, represent a homogeneous and isotropic uneven reflecting surface as a two-dimensional stationary random process $\xi(x,y)$ of changing the height of the irregularities (Figure 3). Any azimuthal cross-section (for example, along the $Y$ axis) of the process $\xi(x,y)$ is a one-dimensional stationary random process $\xi(y)$.

Suppose we know the one-dimensional probability density function (OPDF) $\rho[\xi(y)]$ of the heights of irregularities $\xi(y)$. It will be shown below that on the basis of the OPDF $\rho[\xi(y)]$, it is possible to obtain the OPDF $\rho[\xi'(y)]$ of the derivative of this stationary random process $\xi'(y)$.

Taking into account that $\xi'(y) = \text{tg}\beta$, where $\beta$ is the angle between the tangent to the process $\xi(y)$ and the $Y$ axis (see Figure 4), we make in $\rho[\xi'(y)]$ the change of the variable $\xi'$ to $\beta$. As a result, we obtain the OPDF of angles $\beta$ in the azimuthal section $\xi(y)$ under study

$$\rho(\beta) = \rho(\mathrm{tg}\beta)\left|\frac{1}{\cos^2\beta}\right|, \qquad (2.18)$$

where $|G_\beta| = \left|\dfrac{1}{\cos^2\beta}\right|$ is the Jacobian of the transformation.

Figure 4 shows that between the angles $\beta$ and $\chi$ there is a unambiguous unique functional dependence $\beta + \chi + \pi/2 = \pi$, whence it follows

$$\beta = \pi/2 - \chi. \qquad (2.18)$$

In view of this expression, we make in OPDF $\rho(\beta)$ (2.18) the change of the variable $\beta$ to $\chi$

$$\rho(\beta) = \rho[\mathrm{tg}(\pi/2 - \chi)]\left|\frac{1}{\cos^2(\pi/2 - \chi)}\right|,$$

with the Jacobian of the transformation $|G_\chi| = 1$.

Take into account that $\mathrm{tg}(\pi/2 - \chi) = \mathrm{ctg}\chi$, $\cos(\pi/2 - \chi) = \sin\chi$. As a result, we obtain the OPDF of angles $\chi$

$$\rho(\chi) = \rho(ctg\chi)\left|\frac{1}{\sin^2\chi}\right|, \qquad (2.19)$$

In the case of statistical independence of the angles $\chi$ and $\varphi$ (which is typical for many uneven surfaces), the joint TPDF $\rho(\chi,\varphi)$ (2.1) can be represented as

$$\rho(\chi,\varphi) = \rho(\chi)\rho(\varphi) = \rho(ctg\chi)\left|\frac{1}{\sin^2\chi}\right|\rho(\varphi). \qquad (2.20)$$

For homogeneous and isotropic statistically uneven surfaces, the angle $\varphi$, which determines the azimuthal direction of the projection of the normal to the XOY plane (see Figure 5), can be uniformly distributed in the interval from 0 to $2\pi$, and the OPDF $\rho(\varphi)$ can be given by the expression

$$\rho(\varphi) = \frac{1}{2\pi}. \qquad (2.21)$$

Substituting (2.21) into (2.20), we obtain the TPDF (2.1)

$$\rho(\chi,\varphi) = \frac{1}{2\pi}\left|\frac{1}{\sin^2\chi}\right|\rho(ctg\chi). \qquad (2.22)$$

If the reflecting surface is non-isotropic, then the OPDF $\rho(\varphi)$ can be specified by another function, for example,

$$\rho(\varphi) = \frac{2\sin^2\varphi}{\pi}, \qquad (2.23)$$

or
$$\rho(\varphi) = \frac{4\sin^2\varphi\cos^2\varphi}{\pi}. \qquad (2.23a)$$

In this case, the TPDF (2.1) will have the form

$$\rho(\chi,\varphi) = \frac{2\sin^2\varphi}{\pi}\left|\frac{1}{\sin^2\chi}\right|\rho(ctg\chi) = \frac{2(1-\cos^2\varphi)}{\pi}\left|\frac{1}{\sin^2\chi}\right|\rho(ctg\chi). \qquad (2.24)$$

or $\quad\rho(\chi,\varphi) = \dfrac{4\sin^2\varphi\cos^2\varphi}{\pi}\left|\dfrac{1}{\sin^2\chi}\right|\rho(ctg\chi) = \dfrac{4(1-\cos^2\varphi)\cos^2\varphi}{\pi}\left|\dfrac{1}{\sin^2\chi}\right|\rho(ctg\chi). \qquad (2.24a)$

Note once again that the TPDF $\rho(\chi,\varphi)$ (2.22), (2.24) and (2.24a) are obtained for the case where an uneven surface can be represented as a two-dimensional homogeneous random process of changing the heights of the irregularities $\xi(x,y)$ (see Figure 3). At each point with $x,y$ coordinates, the random variable $\xi$ has the same averaged characteristics: OPDF, expected value, variance, and other moments and central moments.

## 2.5 The overall form of the volumetric diagram of elastic scattering of microparticles

The overall form of volumetric DESM is expressed by (2.4)

$$D(v,\omega/\vartheta,\gamma) = \rho(v,\omega/\vartheta,\gamma)/G_{v\omega}| = \rho\{\chi=f_1(v,\omega/\vartheta,\gamma);\ \varphi=f_2(v,\omega/\vartheta,\gamma)\}|G_{v\omega}|. \qquad (2.25)$$

In view of (2.17) and (2.22), expression (2.25) takes the form

$$D(v,\omega/\vartheta,\gamma) = \frac{1}{2\pi}\left|\frac{1}{\sin^2\{\chi=f_1(v,\omega/\vartheta,\gamma)\}}\right|\rho[ctg\{\chi=f_1(v,\omega/\vartheta,\gamma)\}]\left|\frac{d(a'_v b'_\omega - a'_\omega b'_v) + c'_v(ba'_\omega - ab'_\omega)}{\sqrt{a^2+b^2}(a^2+b^2+d^2)}\right|.$$

We take into account that $1+ctg^2\chi = \dfrac{1}{\sin^2\chi}$, whence follows $ctg\chi = \sqrt{\dfrac{1}{\sin^2\chi}-1}$, therefore, this expression can be represented in the form

$$D(v,\omega/\vartheta,\gamma) = \frac{1}{2\pi}\left|\frac{1}{\sin^2\{\chi=f_1(v,\omega/\vartheta,\gamma)\}}\right|\rho\left(\sqrt{\frac{1}{\sin^2\{\chi=f_1(v,\omega/\vartheta,\gamma)\}}-1}\right)\left|\frac{d(a'_v b'_\omega - a'_\omega b'_v) + c'_v(ba'_\omega - ab'_\omega)}{\sqrt{a^2+b^2}(a^2+b^2+d^2)}\right|. \qquad (2.26)$$

Substituting the functional dependence (2.14) $\chi = \arcsin\left(\dfrac{d}{\sqrt{a^2+b^2+d^2}}\right)$, in (2.26), we obtain

$$D(\nu,\omega/\vartheta,\gamma) = \frac{1}{2\pi}\left|\frac{a^2+b^2+d^2}{d^2}\right|\rho\left(\sqrt{\frac{a^2+b^2}{d^2}}\right)\left|\frac{d(a'_\nu b'_\omega - a'_\omega b'_\nu) + c'_\nu(ba'_\omega - ab'_\omega)}{\sqrt{a^2+b^2}(a^2+b^2+d^2)}\right|.$$

Simplifying this expression, we find the overall form of volumetric DESM (2.25)

$$D(\nu,\omega/\vartheta,\gamma) == \frac{1}{2\pi}\left|\frac{d(a'_\nu b'_\omega - a'_\omega b'_\nu) + c'_\nu(ba'_\omega - ab'_\omega)}{d^2\sqrt{a^2+b^2}}\right|\rho\left(\sqrt{\frac{a^2+b^2}{d^2}}\right), \qquad (2.27)$$

where $a, b, d, a'_\nu, b'_\nu, c'_\nu, a'_\omega, b'_\omega$ are given by expressions (2.11).

Note that formally in expression (2.27), the derivative $\xi'$ was replaced by a quantity $\sqrt{\dfrac{a^2+b^2}{d^2}}$ with the Jacobian of the transformation $|G_{\nu\omega/\vartheta\gamma}| = \left|\dfrac{d(a'_\nu b'_\omega - a'_\omega b'_\nu) + c'_\nu(ba'_\omega - ab'_\omega)}{d^2\sqrt{a^2+b^2}}\right|$.

In the case where the TPDF $\rho(\chi,\varphi)$ has the form (2.24) or (2.24a) (i.e., when the reflecting surface is non-isotopic), then instead of the TPDF (2.27), taking into account (2.15), we obtain

$$D(\nu,\omega/\vartheta,\gamma) = \frac{2}{\pi}\left(\frac{b^2}{a^2+b^2}\right)\left|\frac{d(a'_\nu b'_\omega - a'_\omega b'_\nu) + c'_\nu(ba'_\omega - ab'_\omega)}{d^2\sqrt{a^2+b^2}}\right|\rho\left(\sqrt{\frac{a^2+b^2}{d^2}}\right), \qquad (2.28)$$

or $\quad D(\nu,\omega/\vartheta,\gamma) == \dfrac{4}{\pi}\dfrac{a^2 b^2}{(a^2+b^2)^2}\left|\dfrac{d(a'_\nu b'_\omega - a'_\omega b'_\nu) + c'_\nu(ba'_\omega - ab'_\omega)}{d^2\sqrt{a^2+b^2}}\right|\rho\left(\sqrt{\dfrac{a^2+b^2}{d^2}}\right). \qquad (2.28a)$

Formulas (2.27), (2.28) and (2.28a) can be considered as DESM on a statistically uneven surface under the following conditions (see §2.1):

- the uneven surface is statistically homogeneous;

- the irregularities of this surface are quite smooth and large-scale in comparison with the size of the microparticles; the reflection of microparticles from all local sections of the uneven surface occurs according to the laws of geometric optics (see Figures 4 and 5);

- the portion of the uneven surface involved in the reflection of microparticles is located at a large distance from the generator and detector of microparticles (Figurey 2).

## 2.6 OPDF $\rho[\xi'(r)]$ derivative of the stationary random process $\xi(r)$

In § 2.4 it was shown that in order to determine the DESM on a statistically uneven reflecting surface $\xi(x,y)$, any cross-section of which is described by a stationary random process $\xi(r)$ (Figure 3), it is necessary to find the OPDF $\rho[\xi'(r)]$ derivative of this process.

To search for $\rho[\xi'(r)]$, let's use the method proposed in [34, 35]. If the OPDF $\rho(\xi)$ of the one-dimensional stationary random process (SRP) $\xi(r) = \xi$ is known, then the OPDF $\rho(\xi')$ of the derivative of this process can be obtained on the basis of the following formal procedure [34, 35]:

*a*) The given OPDF $\rho(\xi)$ is represented as the product of two probability amplitudes $\psi(\xi)$:

$$\rho(\xi) = \psi(\xi)\psi(\xi). \tag{2.29}$$

*b*) Two Fourier transforms are performed [34, 35]

$$\psi(\xi') = \frac{1}{\sqrt{2\pi}} \int_{-\infty}^{\infty} \psi(\xi) \exp\{i\xi'\xi/\eta\} d\xi, \tag{2.30}$$

$$\psi*(\xi') = \frac{1}{\sqrt{2\pi}} \int_{-\infty}^{\infty} \psi(\xi) \exp\{-i\xi'\xi/\eta\} d\xi. \tag{2.31}$$

where

$$\eta = \frac{2\sigma_\xi^2}{r_{cor}}, \tag{2.32}$$

$\sigma_\xi$ is the standard deviation of the stationary random process $\xi(r) = \xi$;

$r_{cor}$ is the radius of autocorrelation of this process.

*c*) The desired OPDF of the derivative of the SRP $\xi(r) = \xi$ is [34, 35]:

$$\rho(\xi') = \varphi(\xi')\varphi^*(\xi') = |\varphi(\xi')|^2. \tag{2.33}$$

Let's apply the procedure (2.29) through (2.33) to find the OPDF $\rho[\xi'(r)]$ of the derivative of stationary random process with different statistics of heights of irregularities.

*1] OPDF of the derivative of a Gaussian stationary random process*

Suppose that at each point $r$ of the SRP $\xi(r)$, the random variable $\xi$ (in particular, the height of irregularities) is distributed according to the Gaussian law

$$\rho(\xi) = \frac{1}{\sqrt{2\pi\sigma_{\xi1}^2}} \exp\{-(\xi - a_{\xi1})^2 / 2\sigma_{\xi1}^2\}, \qquad (2.34)$$

where $\sigma_{\xi1}^2$ and $a_{\xi1}$ are the variance and expected value of the given process $\xi(r)$.

According to (2.29), we represent the OPDF (2.34) as the product of two probability amplitudes

$$\rho(\xi) = \psi(\xi)\psi(\xi),$$

where
$$\psi(\xi) = \frac{1}{\sqrt[4]{2\pi\sigma_{\xi1}^2}} \exp\{-(\xi - a_{\xi1})^2 / 4\sigma_{\xi1}^2\}. \qquad (2.35)$$

Let's insert (2.35) into (2.30) and (2.31)

$$\psi(\xi') = \frac{1}{\sqrt{2\pi}} \int_{-\infty}^{\infty} \frac{1}{\sqrt[4]{2\pi\sigma_{\xi1}^2}} \exp\{-(\xi - a_{\xi1})^2 / 4\sigma_{\xi1}^2\}\exp\{i\xi'\xi/\eta\}d\xi, \qquad (2.36)$$

$$\psi*(\xi') = \frac{1}{\sqrt{2\pi}} \int_{-\infty}^{\infty} \frac{1}{\sqrt[4]{2\pi\sigma_{\xi1}^2}} \exp\{-(\xi - a_{\xi1})^2 / 4\sigma_{\xi1}^2\}\exp\{-i\xi'\xi/\eta\}d\xi. \qquad (2.37)$$

Let's perform the integration

$$\psi(\xi') = \frac{1}{\sqrt[4]{2\pi\eta^2/(2\sigma_{\xi1})^2}} \exp\{-\xi'^2/[2\eta/(2\sigma_{\xi1})]^2\}\exp\{ia_{\xi1}\xi'/\eta\}, \qquad (2.38)$$

$$\psi*(\xi') = \frac{1}{\sqrt[4]{2\pi\eta^2/(2\sigma_{\xi1})^2}} \exp\{-\xi'^2/[2\eta/(2\sigma_{\xi1})]^2\}\exp\{-ia_{\xi1}\xi'/\eta\}. \qquad (2.39)$$

In accordance with (2.33), we multiply (2.38) and (2.39), as a result, we obtain

$$\rho(\xi') = \frac{1}{\sqrt{2\pi\sigma_{\xi'}^2}} \exp\{-\xi'^2/2\sigma_{\xi'}^2\}, \qquad (2.40)$$

where according to (2.32) $\qquad \sigma_{\xi'} = \sigma_{\xi1}/r_{\text{cor1}} \qquad (2.41)$

is the standard deviation of the differentiated stationary random process $\xi'(r) = \xi'$;

$r_{cor1}$ is the autocorrelation radius of the initial SRP $\xi(r) = \xi$ with the Gaussian distribution of the height of the irregularities.

2] *OPDF derivative of the SRP with a uniform distribution of the heights of the irregularities*

Suppose that at each point $r$ of the SRP $\xi(r)$, the random variable $\xi$ is distributed according to the uniform law in the interval $\xi_1 < \xi < \xi_2$

$$\rho(\xi) = \frac{1}{\xi_2 - \xi_1}. \tag{2.42}$$

According to (2.29), we represent the OPDF (2.42) as the product of two probability amplitudes

$$\rho(\xi) = \psi(\xi)\psi(\xi), \tag{2.43}$$

where

$$\psi(\xi) = \frac{1}{\sqrt{\xi_2 - \xi_1}}. \tag{2.44}$$

Let's insert (2.43) into (2.30) and (2.31)

$$\psi(\xi') = \frac{1}{\sqrt{2\pi}} \int_{\xi_1}^{\xi_2} \frac{1}{\sqrt{\xi_2 - \xi_1}} \exp\{i\xi'\xi/\eta\}d\xi, \tag{2.45}$$

$$\psi^*(\xi') = \frac{1}{\sqrt{2\pi}} \int_{\xi_1}^{\xi_2} \frac{1}{\sqrt{\xi_2 - \xi_1}} \exp\{-i\xi'\xi/\eta\}d\xi. \tag{2.46}$$

As a result of calculation by the formula (2.45), we obtain

$$\psi(\xi') = \frac{1}{\sqrt{\xi_2 - \xi_1}} \frac{1}{\sqrt{2\pi}} \int_{\xi_1}^{\xi_2} \exp\{i\xi'\xi/\eta\}d\xi = \frac{\exp\{i\xi'\xi_2/\eta\} - \exp\{i\xi'\xi_1/\eta\}}{i\xi'\sqrt{2\pi(\xi_2 - \xi_1)/\eta}}. \tag{2.47}$$

Take into account that $(\xi_2 - \xi_1)/2 = a_{\xi 2}$ is the expected value, $\xi_2 - \xi_1 = l$ is the base of the considered SRP $\xi(r)$. Now we can write $\xi_1 = a_{\xi 2} - l/2$ and $\xi_2 = a_{\xi 2} + l/2$, while the expression (2.47) takes the form

$$\psi(\xi') = \frac{\exp\{i\xi'(a_{\xi 2} + l/2)/\eta\} - \exp\{i\xi'(a_{\xi 2} - l/2)/\eta\}}{i\xi'\sqrt{2\pi l/\eta}} = \frac{\exp\{i\xi'l/(2\eta)\} - \exp\{-i\xi'l/(2\eta)\}}{i\xi'\sqrt{2\pi l/\eta}} \exp\{i\xi'a_{\xi 2}/\eta\}. \tag{2.48}$$

Using the expression $\sin x = \dfrac{e^{ix} - e^{-ix}}{2i}$, we represent (2.48) in the form

$$\psi(\xi') = \frac{2\sin\{\xi' l/(2\eta)\}}{\xi'\sqrt{2\pi l/\eta}} \exp\{i\xi' a_{\xi 2}/\eta\}. \quad (2.49)$$

As a result of similar calculations by the formula (2.46), we obtain

$$\psi^*(\xi') = \frac{2\sin\{\xi' l/(2\eta)\}}{\xi'\sqrt{2\pi l/\eta}} \exp\{-i\xi' a_{\xi 2}/\eta\}. \quad (2.50)$$

Substituting (2.49) and (2.50) in (2.33), we finally find

$$\rho(\xi') = \frac{\sin^2\{\xi' k_2\}}{\xi'^2 \pi k_2}, \quad (2.51)$$

where

$$k_2 = \frac{l}{2\eta} = \frac{3 r_{cor2}}{l} = \frac{3 r_{cor2}}{\xi_2 - \xi_1}, \quad (2.52)$$

$r_{cor\,2}$ is the autocorrelation radius of the initial SRP $\xi(r) = \xi$ with a uniform distribution of the height of the irregularities.

It is taken into account that according to (2.32)

$$\eta = \frac{2\sigma_{\xi 2}^2}{r_{cor2}} = \frac{l^2}{6 r_{cor2}}, \quad (2.53)$$

where $\sigma_{\xi 2}^2 = \frac{l^2}{12} = \frac{(\xi_2 - \xi_1)^2}{12}$ is the dispersion of the SRP $\xi(r) = \xi$ with a uniform distribution of the heights of the irregularities (2.42).

Thus, for an SRP $\xi(r) = \xi$ with a uniform distribution of the heights of the irregularities, OPDF $\rho(\xi')$ its derivative $\xi'$ is a distribution of the type $\sin^2\xi'/\xi'^2$ (2.51) with the scale parameter $k_2$ (2.52).

*3] OPDF derivative of the SRP with the Laplace distribution of the heights of irregularities*

Let at each point $r$ of the SRP $\xi(r) = \xi$ a random variable $\xi$ is distributed according to the Laplace law

$$\rho(\xi) = \frac{1}{2\mu_L} \exp\{-|\xi - a_{\xi 3}|/\mu_L\}, \quad (2.54)$$

where $1/\mu_L$ is the scale parameter of this process $\xi(r) = \xi$;

$a_{\xi 3}$ is the shift parameter (expected value).

According to (2.29), we represent the OPDF (2.54) as the product of two probability amplitudes

$$\rho(\xi) = \psi(\xi)\psi(\xi), \tag{2.55}$$

where
$$\psi(\xi) = \sqrt{\frac{1}{2\mu_L}} \exp\{-|\xi - a_{\xi 3}|/2\mu_L\}. \tag{2.56}$$

Let's insert (2.56) into (2.30) and (2.31)

$$\psi(\xi') = \frac{1}{\sqrt{2\pi}} \int_{-\infty}^{\infty} \sqrt{\frac{1}{2\mu_L}} \exp\{-|\xi - a_{\xi 3}|/2\mu_L\} \exp\{i\xi'\xi/\eta\} d\xi, \tag{2.57}$$

$$\psi*(\xi') = \frac{1}{\sqrt{2\pi}} \int_{-\infty}^{\infty} \sqrt{\frac{1}{2\mu_L}} \exp\{-|\xi - a_{\xi 3}|/2\mu_L\} \exp\{-i\xi'\xi/\eta\} d\xi. \tag{2.58}$$

We rewrite these expressions in the form

$$\psi(\xi') = \frac{2}{\sqrt{4\pi\mu_L}} \int_{a_{\xi 3}}^{\infty} \exp\{-(\xi - a_{\xi 3})/2\mu_L - i\xi'\xi/\eta\} d\xi, \tag{2.59}$$

$$\psi*(\xi') = \frac{2}{\sqrt{4\pi\mu_L}} \int_{-\infty}^{a_{\xi 3}} \exp\{-(\xi - a_{\xi 3})/2\mu_L + i\xi'\xi/\eta\} d\xi. \tag{2.60}$$

Let's perform the integration

$$\psi(\xi') = \sqrt{\frac{\mu_L \eta}{\pi}} \frac{\exp(a_{\xi 3}/2\mu_L - a_{\xi 3})}{\eta/2 + i\xi'\mu_L}, \tag{2.61}$$

$$\psi*(\xi') = \sqrt{\frac{\mu_L \eta}{\pi}} \frac{\exp(-a_{\xi 3}/2\mu_L + a_{\xi 3})}{\eta/2 - i\xi'\mu_L}. \tag{2.62}$$

Substituting (2.61) and (2.62) in (2.33), we find

$$\rho(\xi') = \frac{\mu_L \eta}{\pi(\eta^2/4 + \xi'^2\mu_L^2)}. \tag{2.63}$$

The dispersion of the Laplace distribution (2.54) is $\sigma_{\xi 3}^2 = 2\mu_L^2$, therefore, according to (2.32), in this case

$$\eta = \frac{4\mu_L^2}{r_{cor3}}, \tag{2.64}$$

where $r_{cor3}$ is the autocorrelation radius of the initial SRP $\xi(r) = \xi$ with the Laplace distribution of the height of irregularities.

Substituting (2.64) in (2.63), we obtain

$$\rho(\xi') = \frac{k_3}{\pi(k_3^2 + \xi'^2)}, \tag{2.65}$$

where $k_3 = \dfrac{2\mu_L}{r_{cor3}}$ is the scale parameter.

Thus, for the Laplace SRP $\xi(r) = \xi$ of the OPDF $\rho(\xi')$ its derivative $\xi'$ is the Cauchy distribution (2.65) with the expected value (i.e., the shift parameter) equal to zero.

*4] OPDF derivative of the SRP with the distribution of the height of the irregularities according to the Cauchy law*

Let at each point $r$ of the SRP $\xi(r) = \xi$ a random variable $\xi$ be distributed according to the Cauchy law

$$\rho(\xi) = \frac{\mu_K}{\pi[\mu_K^2 + (\xi - a_{\xi 4})^2]}, \tag{2.66}$$

where $\mu_K$ is the scale parameter of this process $\xi(r) = \xi$;

$a_{\xi 4}$ is the shift parameter (expected value).

Performing actions (2.29) through (2.33), the reverse of transformations (2.57) through (2.63), we obtain

$$\rho(\xi') = \frac{1}{2k_4}\exp\{-|\xi'|/k_4\}. \tag{2.67}$$

To find the scale parameter $k_4$, we note that the dispersion (variance) of the Cauchy distribution is not defined, i.e., tends to infinity, but the heights of the real surface irregularities can be distributed only by a truncated Cauchy law with effective dispersion $\sigma_{\xi 4}^2 \sim 25\mu_K^2$. Therefore, in this case, according to (2.32), we can write

$$\eta \approx \frac{50\mu_K^2}{r_{cor4}}, \tag{2.68}$$

where $r_{cor4}$ is the autocorrelation radius of the initial SRP $\xi(r) = \xi$ with the distribution of the heights of irregularities $\xi$ according to the Cauchy law (2.66).

In this case, we obtain the following estimate of the scale parameter

$$k_4 \approx 2\mu_K / \eta \approx \frac{r_{cor4}}{25\mu_K}. \qquad (2.69)$$

Thus, for an SRP with a distribution of roughness heights $\xi(r) = \xi$ according to the Cauchy law (2.66), the OPDF $\rho(\xi')$ of its derivative $\xi'$ is the Laplace distribution (2.67) with the scale parameter (2.69) and the shift parameter (expected value) equal to zero.

*5] OPDF derivative of the SRP with the distribution of the heights of the irregularities according to the multilayer sinusoidal law*

Consider the scattering of microparticles (in particular, electrons or high-frequency photons) on a single crystal. Falling microparticles can be reflected from different atomic layers of the crystal lattice of a single crystal (Figure 6*a*). This case is equivalent to the scattering of microparticles on a multilayer reflecting surface, each layer of which can be defined by a two-dimensional homogeneous SRP $\xi(x,y)$, repeating on average the structure of a planar atomic lattice (Figure 6*b*).

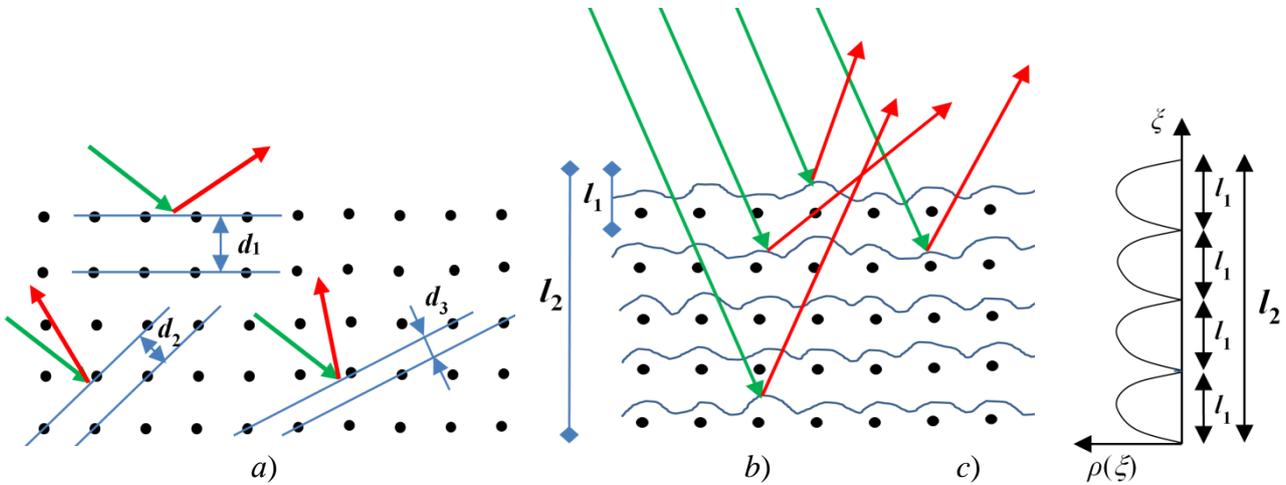

**F**ig. 6 *a*) Reflection of falling microparticles from various atomic layers of the crystal lattice of a single crystal; *b*) Scattering of microparticles on the multilayer surface of the crystal, with each layer being considered as a separate uneven surface of the sinusoidal type; *c*) Multi-humped sinusoidal OPDF of the height of the irregularities of the multilayer surface of the crystal (2.71)

If only the upper layer of the crystal is involved in the scattering of microparticles, then it can be assumed that in each azimuthal cross-section *r* of such an SRP $\xi(r) = \xi$, a random value $\xi$ (i.e., the height of the unevenness of the upper layer of the crystal surface) is distributed according to the sinusoidal law

$$\rho(\xi) = \begin{cases} \dfrac{2\sin^2(\pi\xi/l_1)}{l_1} & npu\ \xi \in [0, l_1]; \\ 0 & npu\ \xi \notin [0, l_1], \end{cases} \quad (2.70)$$

where $l_1$ is the thickness of one (i.e., the first) reflecting layer of a single crystal (Figure 6c).

In the case where several identical crystal layers are effectively involved in the scattering of microparticles (Figure 6b), then the multi-humped sinusoidal OPDF of the heights of irregularities (Figure 6c) should be used.

$$\rho(\xi) = \begin{cases} \dfrac{2\sin^2(\pi n_1 \xi/l_2)}{l_2} & npu\ \xi \in [0, l_2]; \\ 0 & npu\ \xi \notin [0, l_2], \end{cases} \quad (2.71)$$

where $n_1$ is the number of identical uneven sinusoidal layers lying in the interval $[0, l_2]$, here $l_2 = n_1 l_1$ is the depth of the multilayer crystal surface of the effectively scattering microparticles.

According to (2.29), we represent the OPDF (2.71) as the product of two probability amplitudes

$$\rho(\xi) = \psi(\xi)\psi(\xi),$$

where
$$\psi(\xi) = \sqrt{\dfrac{2}{l_2}} \sin(\pi n_1 \xi/l_2).$$

Let's insert (2.72) into (2.30) and (2.31)

$$\psi*(\xi') = \dfrac{1}{\sqrt{2\pi}} \int_0^{l_2} \sqrt{\dfrac{2}{l_2}} \sin(\pi n_1 \xi/l_2) \exp\{i\xi'\xi/\eta\} d\xi, \quad (2.73)$$

$$\psi*(\xi') = \dfrac{1}{\sqrt{2\pi}} \int_0^{l_2} \sqrt{\dfrac{2}{l_2}} \sin(\pi n_1 \xi/l_2) \exp\{-i\xi'\xi/\eta\} d\xi.$$

Performing integration [see Appendix 2, (A.2.12) and (A.2.13)], we obtain

$$\psi(\xi') = -\sqrt{\dfrac{1}{4\pi l_2}} \left( \dfrac{e^{i(\pi n_1 + \xi' l_2/\eta)} - 1}{(\pi n_1/l_2 + \xi'/\eta)} + \dfrac{e^{-i(\pi n_1 - \xi' l_2/\eta)} - 1}{(\pi n_1/l_2 - \xi'/\eta)} \right), \quad (2.74)$$

$$\psi^*(\xi') = -\sqrt{\frac{1}{4\pi l_2}} \left( \frac{e^{i(\pi n_1 - \xi' l_2/\eta)} - 1}{(\pi n_1/l_2 - \xi'/\eta)} + \frac{e^{-i(\pi n_1 + \xi' l_2/\eta)} - 1}{(\pi n_1/l_2 + \xi'/\eta)} \right). \quad (2.75)$$

Substituting (2.74) and (2.75) into (2.33) [see Appendix 3, (A.3.12)], we find

$$p(\xi') = \psi(\xi')\psi^*(\xi') = \frac{1}{\pi l_2} \left( \frac{\cos^2(\pi n_1) - \cos(\pi n_1)\cos(\xi' l_2/\eta)}{\left(\pi n_1/l_2\right)^2 - \left(\xi'/\eta\right)^2} - \frac{\cos(\pi n_1 + \xi' l_2/\eta) - 1}{\left(\pi n_1/l_2 + \xi'/\eta\right)^2} \right). \quad (2.76)$$

The dispersion (variance) of the multi-humped sinusoidal distribution (2.71) is equal to

$$\sigma_{\xi 4}^2 = \frac{l_2^2(\pi^2 n_1^2 - 6)}{12\pi^2 n_1^2} = \frac{l_1^2 n_1^2 (\pi^2 n_1^2 - 6)}{12\pi^2 n_1^2} = \frac{l_1^2 (\pi^2 n_1^2 - 6)}{12\pi^2}. \quad (2.77)$$

Therefore, in this case, according to (2.32), we have the scale parameter

$$\eta = \frac{2\sigma_{\xi 41}^2}{r_{cor5}} = \frac{l_1^2 (\pi^2 n_1^2 - 6)}{6\pi^2 r_{cor5}}, \quad (2.78)$$

where $r_{cor5}$ is the autocorrelation radius of one uneven layer of a sinusoidal crystal.

OPDF (2.76) can be represented in the form

$$p(\xi') = \frac{1}{\pi l_2} \left( \frac{\cos(\pi n_1)[\cos(\pi n_1) - \cos(\xi' l_2/\eta)]}{\left(\pi n_1/l_2\right)^2 - \left(\xi'/\eta\right)^2} - \frac{\cos(\pi n_1 + \xi' l_2/\eta) - 1}{\left(\pi n_1/l_2 + \xi'/\eta\right)^2} \right). \quad (2.79)$$

Taking into account the trigonometric formula $\cos x - \cos y = 2\sin\left(\frac{y+x}{2}\right)\sin\left(\frac{y-x}{2}\right)$ from expression (2.79), we obtain another form of the desired OPDF

$$p(\xi') = \frac{1}{\pi l_2} \left( \frac{2\cos(\pi n_1)\sin\left(\frac{\xi' l_2/\eta + \pi n_1}{2}\right)\sin\left(\frac{\xi' l_2/\eta - \pi n_1}{2}\right)}{\left(\pi n_1/l_2\right)^2 - \left(\xi'/\eta\right)^2} - \frac{\cos(\pi n_1 + \xi' l_2/\eta) - 1}{\left(\pi n_1/l_2 + \xi'/\eta\right)^2} \right). \quad (2.80)$$

Thus, for an SRP with a multi-humped sinusoidal OPDF of the heights of irregularities (2.71) of the OPDF $\rho(\xi')$ its derivative $\xi'$ is distribution (2.76) [or in another form (2.80)] with scale parameter (2.78).

Using the formal procedure (2.29) through (2.33), we can obtain the OPDF $\rho(\xi')$ of the derivative $\xi'$ for many other stationary random processes with different statistics of the height of irregularities $\xi$.

OPDF $\rho(\xi')$ (2.40), (2.51), (2.65), (2.67), (2 of irregularities.76) and others can be used in many problems of static physics.

## 3 Results

Based on the method proposed in Section 2, in this part of the paper we obtain formulas for calculating volumetric diagrams of elastic scattering of microparticles (DESM) on statistically uneven surfaces when the conditions of the Kirchhoff approximation are met.

### 3.1 Volumetric diagrams of elastic scattering of microparticles on statistically uneven surfaces

*1] Volumetric diagram of elastic scattering of microparticles on a reflecting surface with a Gaussian distribution of the heights of irregularities*

As an example, we consider the procedure for obtaining a volumetric DESM $D(v,\omega/\vartheta,\gamma)$ (2.27) for the case where the homogeneous and isotropic irregularities $\xi(x,y)$ of the reflecting surface (see Figure 3) at each point with coordinates $x,y$ are distributed according to the law of Gauss (2.34).

In this case, the OPDF $\rho(\xi')$ of the derivative of this stationary random process is also Gaussian (2.40)

$$\rho(\xi') = \frac{1}{\sqrt{2\pi\sigma_{\xi'}^2}} \exp\left(-\xi'^2/2\sigma_{\xi'}^2\right), \tag{3.1}$$

where $\sigma_{\xi'} = \sigma_{\xi 1}/r_{\text{cor}1}$.

In accordance with the algorithm described at the end of § 2.5, instead of $\xi'$ in (3.1), we substitute $\sqrt{\dfrac{a^2+b^2}{d^2}}$, as a result, we obtain

$$\rho\left(\sqrt{\frac{a^2+b^2}{d^2}}\right) = \frac{1}{\sqrt{2\pi\sigma_{\xi'}^2}} \exp\left(-\frac{a^2+b^2}{2d^2\sigma_{\xi'}^2}\right). \tag{3.2}$$

Substituting (3.2) into (2.27), we obtain the explicit form of the desired DESM in the case of a Gaussian distribution of the heights of the irregularities of the reflecting surface

$$D(v,\omega/\vartheta,\gamma) = \frac{1}{\sqrt{8\pi^3\sigma_{\xi'}^2}} \exp\left(-\frac{a^2+b^2}{2d^2\sigma_{\xi'}^2}\right) \left|\frac{d(a'_v b'_\omega - a'_\omega b'_v) + c'_v(ba'_\omega - ab'_\omega)}{d^2\sqrt{a^2+b^2}}\right|, \quad (3.3)$$

where $\sigma_{\xi'} = \sigma_{\xi 1}/r_{cor1}$; the values of $a, b, d, a'_v, b'_v, c'_v, a'_\omega, b'_\omega$ are given by expressions (2.11).

Under the conditions specified at the end of §2.5, the expression (3.3) is a formula for calculating volumetric DESM on a large-scale (as compared to the size of microparticles) uneven reflecting surface with a Gaussian distribution of the heights of the irregularities $\xi(x,y)$.

The scattering diagrams calculated by formula (3.3) for various values of $\vartheta$, $\gamma$, $\sigma_{\xi 1}$ and $r_{cor1}$ are shown in Figure 7 (see Appendix 4).

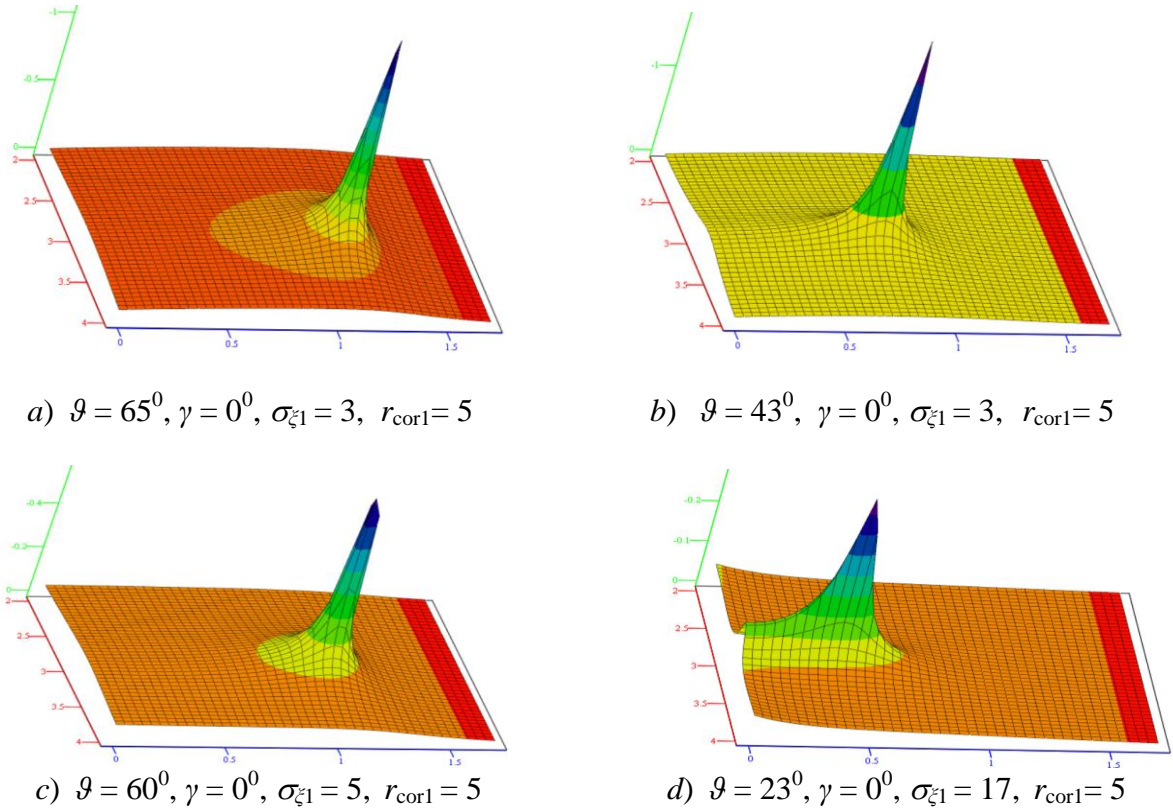

a) $\vartheta = 65^0$, $\gamma = 0^0$, $\sigma_{\xi 1} = 3$, $r_{cor1} = 5$     b) $\vartheta = 43^0$, $\gamma = 0^0$, $\sigma_{\xi 1} = 3$, $r_{cor1} = 5$

c) $\vartheta = 60^0$, $\gamma = 0^0$, $\sigma_{\xi 1} = 5$, $r_{cor1} = 5$     d) $\vartheta = 23^0$, $\gamma = 0^0$, $\sigma_{\xi 1} = 17$, $r_{cor1} = 5$

**Fig. 7** Volumetric DESM on a homogeneous and isotropic uneven surface with a Gaussian distribution of the heights of irregularities. The calculations were performed according to the formula (3.3) for various values of the parameters $\vartheta$, $\gamma$, $\sigma_{\xi 1}$ and $r_{cor1}$. Here and further, DESM are calculated using MathCad software

*2] Volumetric DESM on a reflecting surface with a uniform distribution of the heights of irregularities*

Let the homogeneous and isotropic irregularities $\xi(x,y)$ of the reflecting surface, at each point with coordinates $x,y$ be distributed according to the uniform law (2.42). In this case, we use the OPDF $\rho(\xi')$ (2.51). Performing actions similar to (3.1) through (3.3), we obtain a formula for calculating volumetric DESM on a given surface

$$D(\nu,\omega/\vartheta,\gamma) = \frac{1}{2\pi^2 k_2}\sin^2\left(k_2\sqrt{\frac{a^2+b^2}{d^2}}\right)\left|\frac{d(a'_\nu b'_\omega - a'_\omega b'_\nu) + c'_\nu(ba'_\omega - ab'_\omega)}{(a^2+b^2)^{3/2}}\right|, \qquad (3.4)$$

where $k_2 = \dfrac{3r_{cor2}}{l} = \dfrac{3r_{cor2}}{\xi_2 - \xi_1}$ is the scale parameter.

The results of calculations using the formula (3.4) for different values of the parameters $\vartheta$, $\gamma$, $l$ and $r_{cor2}$ are shown in Fig. 8 (see Appendix 5).

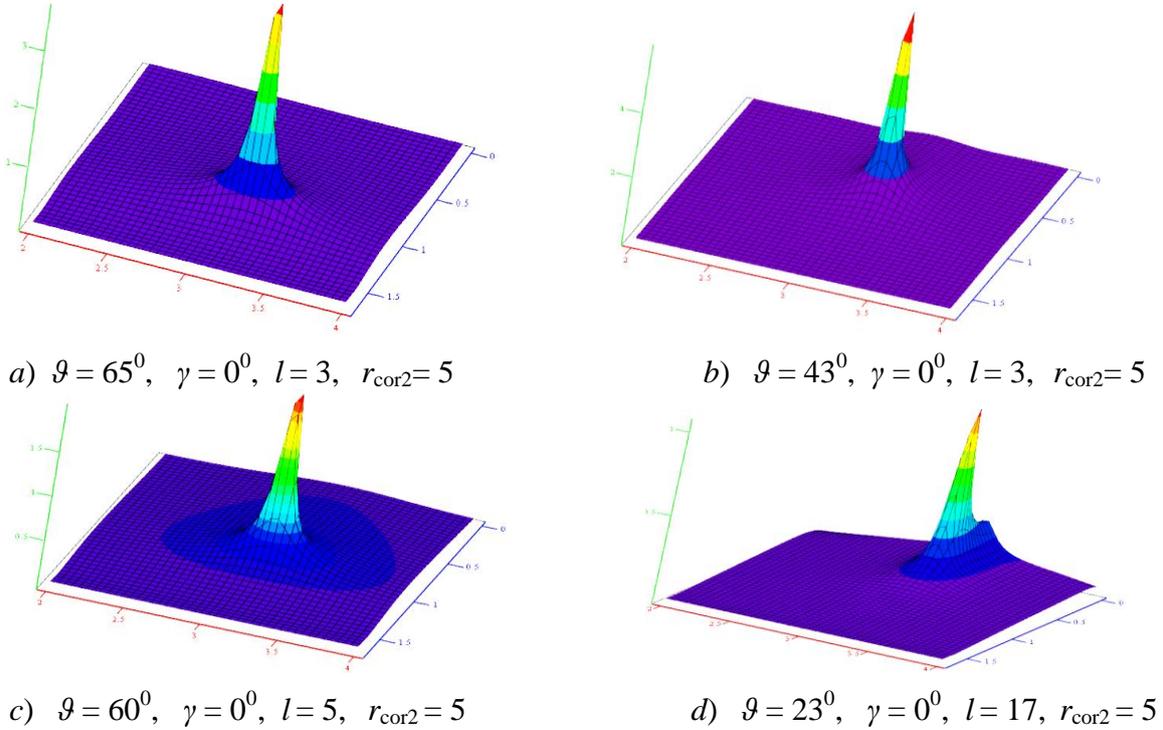

a) $\vartheta = 65^0$, $\gamma = 0^0$, $l = 3$, $r_{cor2} = 5$   b) $\vartheta = 43^0$, $\gamma = 0^0$, $l = 3$, $r_{cor2} = 5$

c) $\vartheta = 60^0$, $\gamma = 0^0$, $l = 5$, $r_{cor2} = 5$   d) $\vartheta = 23^0$, $\gamma = 0^0$, $l = 17$, $r_{cor2} = 5$

**Fig. 8** Volumetric DESM on a homogeneous and isotropic uneven reflecting surface with a uniform distribution of the heights of irregularities. The calculations are performed according to the formula (3.4)

*2] Volumetric DESM on a reflecting surface with a Laplace distribution of the heights of irregularities*

Let the homogeneous and isotropic irregularities $\xi(x,y)$ of the reflecting surface, at each point with coordinates $x,y$ be distributed according to the Laplace law (2.54). In this case, we use the OPDF $\rho(\xi')$ (2.65). Performing actions similar to (3.1) through (3.3), we obtain a formula for calculating volumetric DESM on a given surface

$$D(\nu,\omega/\vartheta,\gamma) = \frac{k_3}{2\pi^2(k_3^2 d^2 + a^2 + b^2)} \left| \frac{d(a'_\nu b'_\omega - a'_\omega b'_\nu) + c'_\nu(ba'_\omega - ab'_\omega)}{\sqrt{a^2 + b^2}} \right|, \qquad (3.5)$$

where $k_3 = \dfrac{2\mu_L}{r_{cor3}}$ is the scale parameter.

The results of calculations using the formula (3.5) for different values of the parameters $\vartheta$, $\gamma$, $\mu_L$ and $r_{cor3}$ are shown in Figure 9 (see Appendix 6).

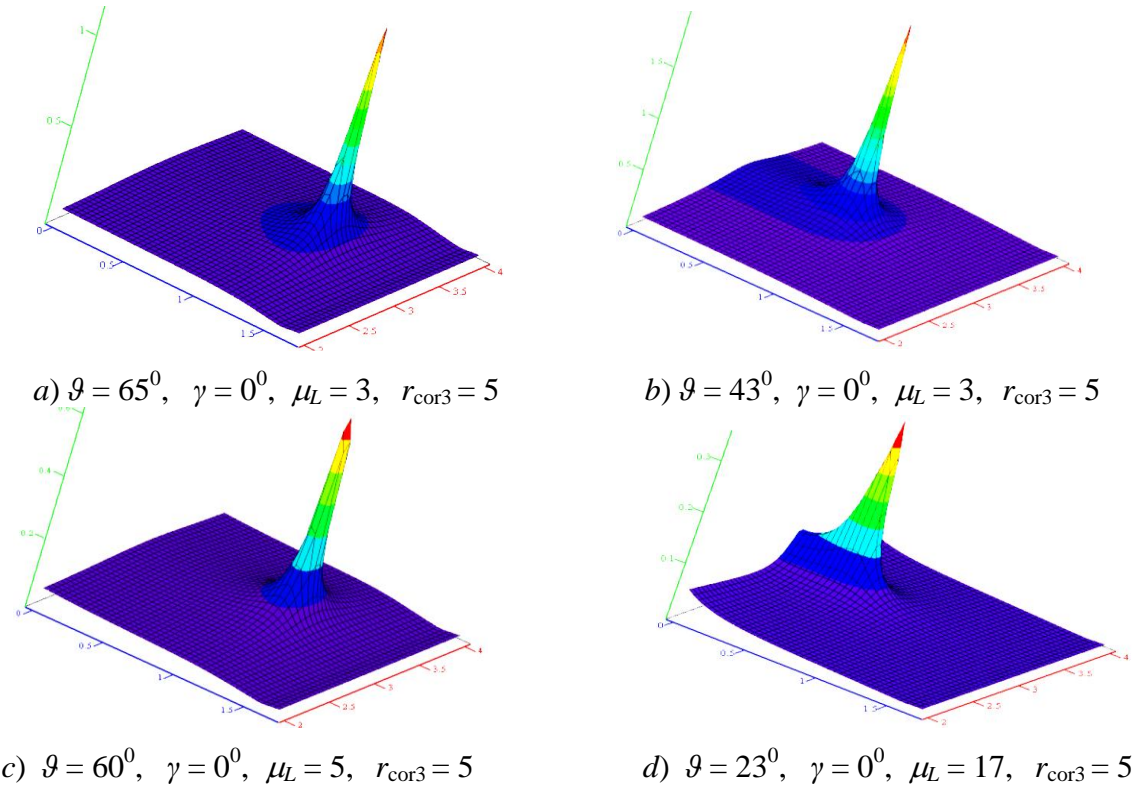

a) $\vartheta = 65^0$, $\gamma = 0^0$, $\mu_L = 3$, $r_{cor3} = 5$      b) $\vartheta = 43^0$, $\gamma = 0^0$, $\mu_L = 3$, $r_{cor3} = 5$

c) $\vartheta = 60^0$, $\gamma = 0^0$, $\mu_L = 5$, $r_{cor3} = 5$      d) $\vartheta = 23^0$, $\gamma = 0^0$, $\mu_L = 17$, $r_{cor3} = 5$

**Fig. 9** Volumetric DESM on a homogeneous and isotropic uneven surface with a Laplace distribution of the heights of irregularities. The calculations are performed according to the formula (3.5)

*4] Volumetric DESM on a reflecting surface with a distribution of the heights of irregularities according to the Cauchy law*

Let the homogeneous and isotropic irregularities $\xi(x,y)$ of the reflecting surface, at each point with coordinates $x,y$ be distributed according to the Cauchy law (2.66). In this case, we use the OPDF $\rho(\xi')$ (2.67). Performing actions similar to (3.1) through (3.3), we obtain a formula for calculating volumetric DESM on a given surface

$$D(\nu,\omega/\vartheta,\gamma) = \frac{1}{2k_4^2} \exp\left(-\sqrt{\frac{a^2+b^2}{d^2 k_4^2}}\right) \left| \frac{d(a_\nu' b_\omega' - a_\omega' b_\nu') + c_\nu'(ba_\omega' - ab_\omega')}{d^2 \sqrt{a^2+b^2}} \right|, \quad (3.6)$$

where $k_4 \approx \dfrac{r_{cor4}}{25\mu_K}$ is the scale parameter.

The results of calculations using the formula (3.6) for different values of the parameters $\vartheta$, $\gamma$, $\mu_K$ and $r_{cor4}$ are shown in Figure 9 (see Appendix 7).

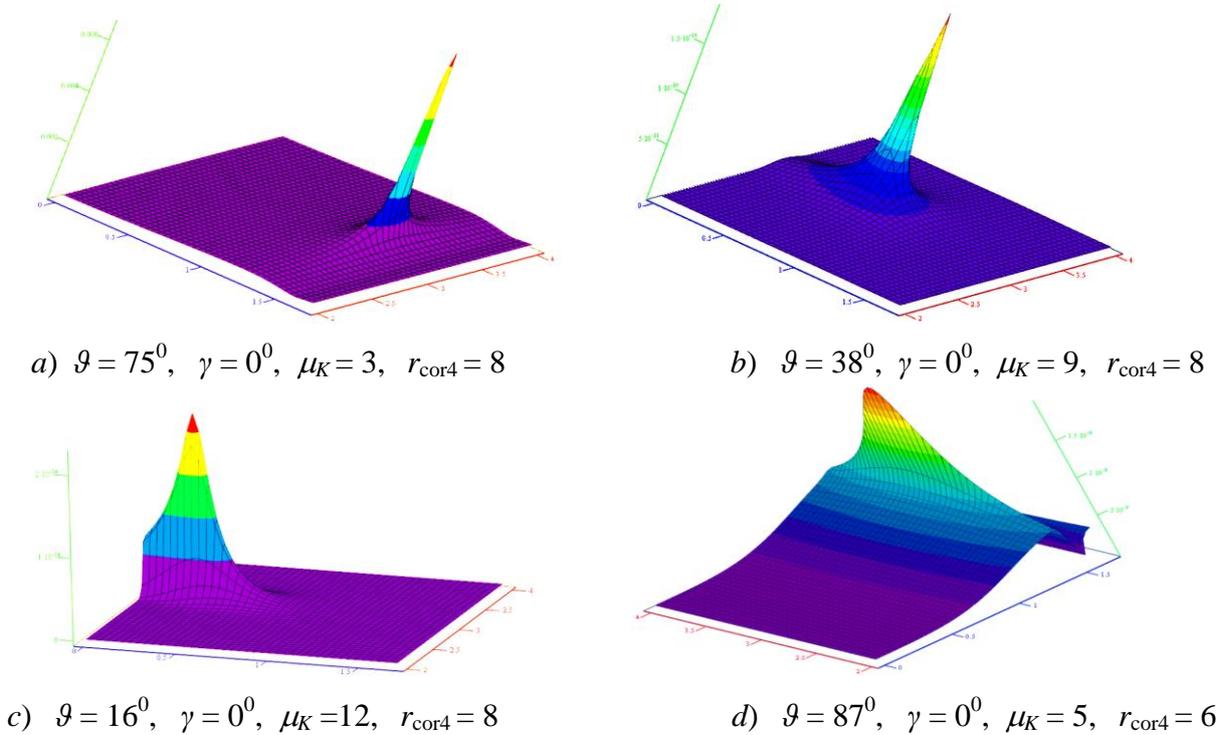

a) $\vartheta = 75^0$, $\gamma = 0^0$, $\mu_K = 3$, $r_{cor4} = 8$  b) $\vartheta = 38^0$, $\gamma = 0^0$, $\mu_K = 9$, $r_{cor4} = 8$

c) $\vartheta = 16^0$, $\gamma = 0^0$, $\mu_K = 12$, $r_{cor4} = 8$  d) $\vartheta = 87^0$, $\gamma = 0^0$, $\mu_K = 5$, $r_{cor4} = 6$

**Fig. 10** Volumetric DESM on a homogeneous and isotropic uneven surface with distribution of the heights of irregularities according to the truncated Cauchy law. The calculations are performed according to the formula (3.6)

Analysis of the scattering diagrams shown in Figures 7 through 10, as well as of other DESMs calculated by the formulas (3.3) through (3.6), shows that the elastic scattering of microparticles on one upper layer of a statistically uneven reflecting surface weakly depends on the statistics of the heights of its irregularities.

Single-layer surfaces with a distribution of the heights of irregularities according to the Gauss law, the uniform law, the Laplace law, and the Cauchy law scatter microparticles almost equally. Some differences between DESMs shown in Figures 7 through 10 are observed at small slip angles of incident microparticles $\vartheta$ and large values of the ratio $\sigma_{\xi m} / r_{\text{cor}}$.

## 3.2 Volumetric diagram of elastic scattering of microparticles on a multilayer surface of a crystal with a multi-humped sinusoidal distribution of heights of irregularities

Let's consider the scattering of microparticles on an $n_1$-layer surface of a crystal, each layer of which is a homogeneous and isotropic two-dimensional uneven surface $\xi(x,y)$ with sinusoidal irregularities (Figure 6b). In this case, we use the multi-humped sinusoidal distribution of the height of the irregularities $\xi$ (2.71), while the OPDF of $\rho(\xi')$ its derivative is expression (2.76) with the scale parameter $\eta$ (2.78)

$$p(\xi') = \frac{1}{\pi l_2} \left( \frac{\cos^2(\pi n_1) - \cos(\pi n_1)\cos(\xi' l_2 / \eta)}{\left(\pi n_1 / l_2\right)^2 - \left(\xi' / \eta\right)^2} - \frac{\cos(\pi n_1 + \xi' l_2 / \eta) - 1}{\left(\pi n_1 / l_2 + \xi' / \eta\right)^2} \right), \qquad (3.7)$$

or the same function in the form (2.80)

$$p(\xi') = \frac{1}{\pi l_2} \left( \frac{2\cos(\pi n_1)\sin\left(\frac{\xi' l_2}{2\eta} + \frac{\pi n_1}{2}\right)\sin\left(\frac{\xi' l_2}{2\eta} - \frac{\pi n_1}{2}\right)}{\left(\pi n_1 / l_2\right)^2 - \left(\xi' / \eta\right)^2} - \frac{\cos(\pi n_1 + \xi' l_2 / \eta) - 1}{\left(\pi n_1 / l_2 + \xi' / \eta\right)^2} \right). \qquad (3.8)$$

To obtain a volumetric DESM on a multilayer crystal surface, we use the method described in § 2.5 [similar to (3.1) through (3.3)]. Replace the derivative $\xi'$ in (3.7) [or in (3.8)] by the value $\sqrt{\dfrac{a^2+b^2}{d^2}}$, and substitute the resulting expression in (2.27). As a result, we obtain the formula for calculating the DESM on a multilayer crystal surface

$$D(\nu,\omega/\vartheta,\gamma) = \frac{1}{2\pi^2 l_2}\left(\frac{\cos^2(\pi n_1)-\cos(\pi n_1)\cos\left(\sqrt{\dfrac{a^2+b^2}{d^2}}l_2/\eta\right)}{\left(\pi n_1/l_2\right)^2 - \left(\sqrt{\dfrac{a^2+b^2}{d^2}}l_2/\eta\right)^2} - \frac{\cos\left(\pi n_1+\sqrt{\dfrac{a^2+b^2}{d^2}}l_2/\eta\right)-1}{\left(\pi n_1/l_2 + \sqrt{\dfrac{a^2+b^2}{d^2}}/\eta\right)^2}\right) \times$$

$$\times \left|\frac{d(a'_\nu b'_\omega - a'_\omega b'_\nu) + c'_\nu(ba'_\omega - ab'_\omega)}{d^2\sqrt{a^2+b^2}}\right|, \tag{3.9}$$

or the same formula in another form

$$D(\nu,\omega/\vartheta,\gamma) = \frac{1}{2\pi^2 l_2}\left(\frac{2\cos(\pi n_1)\sin\left(\sqrt{\dfrac{a^2+b^2}{d^2}}\dfrac{l_2}{2\eta}+\dfrac{\pi n_1}{2}\right)\sin\left(\sqrt{\dfrac{a^2+b^2}{d^2}}\dfrac{l_2}{2\eta}-\dfrac{\pi n_1}{2}\right)}{\left(\pi n_1/l_2\right)^2 - \left(\sqrt{\dfrac{a^2+b^2}{d^2}}l_2/\eta\right)^2} - \frac{\cos\left(\pi n_1+\sqrt{\dfrac{a^2+b^2}{d^2}}l_2/\eta\right)-1}{\left(\pi n_1/l_2 + \sqrt{\dfrac{a^2+b^2}{d^2}}/\eta\right)^2}\right) \times$$

$$\times \left|\frac{d(a'_\nu b'_\omega - a'_\omega b'_\nu) + c'_\nu(ba'_\omega - ab'_\omega)}{d^2\sqrt{a^2+b^2}}\right|,$$

(3.10)

where according to (2.11)

$a = \cos\nu\cos\omega + \cos\vartheta\cos\gamma;\quad b = \cos\nu\sin\omega + \cos\vartheta\sin\gamma;\quad d = \sin\nu + \sin\vartheta;\quad a'_\nu = -\sin\nu\cos\omega;$

$b'_\nu = -\sin\nu\sin\omega;\quad c'_\nu = \cos\nu;\quad a'_\omega = -\cos\nu\sin\omega;\quad b'_\omega = \cos\nu\cos\omega\,;$

according to (2.78)

$$\eta = \frac{l_1^2(\pi^2 n_1^2 - 6)}{6\pi^2 r_{cor5}}, \tag{3.11}$$

where

$l_1$ is the thickness of one reflecting layer (i.e., the horizontal atomic plane) of the crystal (Fig. 6*b*);

$l_2 = l_1 n_1$ is the depth of the multilayer surface of the single crystal, effectively involved in the elastic scattering of microparticles (s);

$n_1$ is the number of uneven layers of a single-crystal (sinusoidal type) that fit in the interval $[0, l_2]$;

$r_{cor5}$ is the autocorrelation radius of one uneven layer of a sinusoidal type. This autocorrelation radius is approximately equal to the average radius of curvature of the sinusoidal irregularities of a single crystal layer;

$\vartheta, \gamma$ are the angles that specify the direction of motion of the microparticle beam incident on the crystal surface (Figures 2, 3, 5);

$v, \omega$ are the angles that specify the direction of movement of microparticles reflected from the surface of the crystal toward the detector (Figures 2, 3, 5).

Expressions (3.9) and (3.10) are the same formula for calculating DESM on a multilayer crystal surface, only written in different trigonometric forms. The formula (3.9) will be called the cosine-version of the DESM on the multilayer resulting surface, and the formula (3.10) is the sinus-version of the same DESM.

The DESM calculated by the formula (3.9) for various values of the five parameters $\vartheta$, $\gamma$, $l_1$, $n_1$ and $r_{cor5}$ are shown in Figure 11 (see Appendix 8).

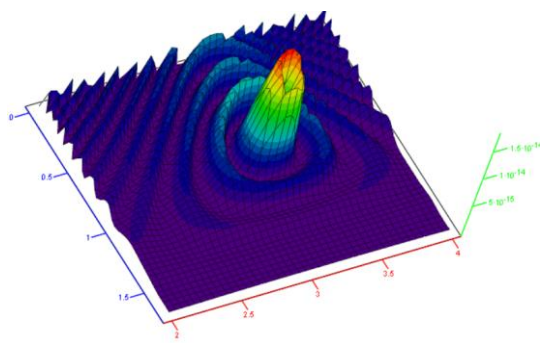
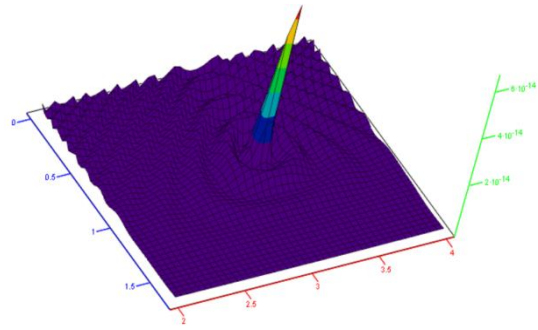

a) $\vartheta = 45^0$, $\gamma = 0^0$, $n_1 = 64$,
$l_1 = 10^{-11}$cm, $r_{cor5} = 6 \times 10^{-9}$cm

b) $\vartheta = 45^0$, $\gamma = 0^0$, $n_1 = 65$,
$l_1 = 10^{-11}$cm, $r_{cor5} = 6 \times 10^{-9}$cm

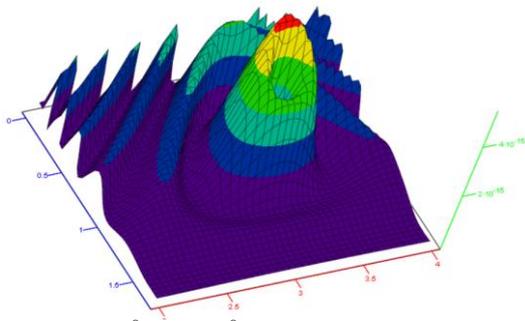
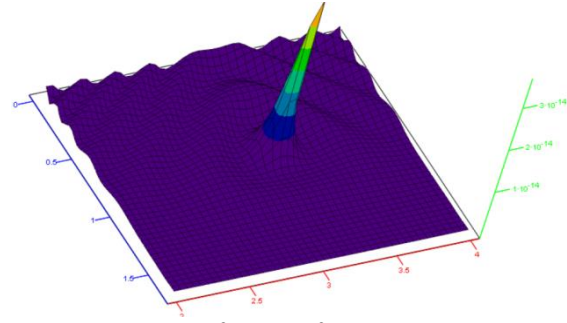

c) $\vartheta = 45^0$, $\gamma = 0^0$, $n_1 = 126$, $l_1 = 10^{-11}$cm, $r_{cor5} = 6\times 10^{-9}$cm

d) $\vartheta = 45^0$, $\gamma = 0^0$, $n_1 = 127$, $l_1 = 10^{-11}$cm, $r_{cor5} = 6\times 10^{-9}$cm

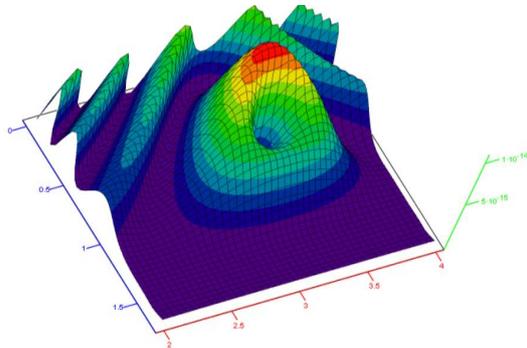
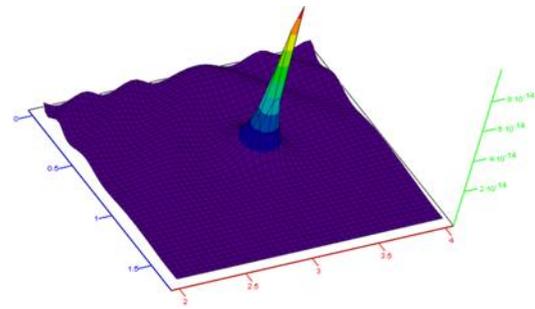

e) $\vartheta = 45^0$, $\gamma = 0^0$, $n_1 = 46$, $l_1 = 10^{-11}$cm, $r_{cor5} = 1{,}4\times 10^{-9}$cm

f) $\vartheta = 45^0$, $\gamma = 0^0$, $n_1 = 47$, $l_1 = 10^{-11}$cm, $r_{cor5} = 1{,}4\times 10^{-9}$cm

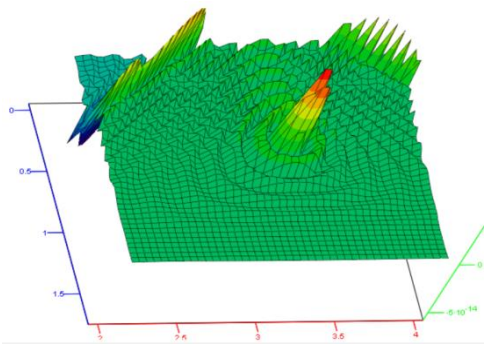
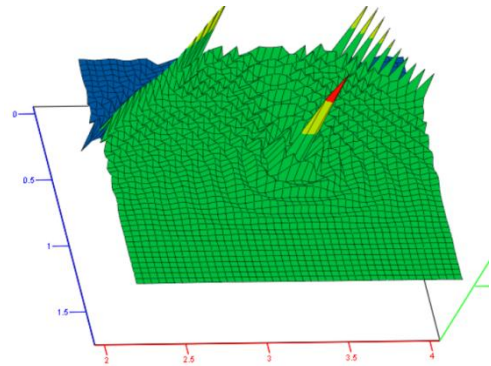

g) $\vartheta = 45^0$, $\gamma = 0^0$, $n_1 = 24$, $l_1 = 10^{-11}$cm, $r_{cor5} = 4\times 10^{-9}$cm

h) $\vartheta = 45^0$, $\gamma = 0^0$, $n_1 = 23$, $l_1 = 10^{-11}$cm, $r_{cor5} = 4\times 10^{-9}$cm

**Fig. 11** Volumetric diagrams of elastic scattering of microparticles on a multilayer crystal surface, calculated by the formula (3.9) for various values of the parameters $\vartheta$, $l_1$, $n_1$ and $r_{cor5}$

If each crystal layer has the same anisotropy, for example, of type (2.23), then, taking into account (2.24) and (2.28), we obtain the following formula for calculating the volumetric DESM for this case

$$D(\nu,\omega/\vartheta,\gamma) = \frac{2}{\pi^2 l_2}\left(\frac{b^2}{a^2+b^2}\right)\left(\frac{\cos^2(\pi n_1) - \cos(\pi n_1)\cos\left(\sqrt{\frac{a^2+b^2}{d^2}}l_2/\eta\right)}{\left(\pi n_1/l_2\right)^2 - \left(\sqrt{\frac{a^2+b^2}{d^2}}l_2/\eta\right)^2} - \frac{\cos\left(\pi n_1 + \sqrt{\frac{a^2+b^2}{d^2}}l_2/\eta\right) - 1}{\left(\pi n_1/l_2 + \sqrt{\frac{a^2+b^2}{d^2}}/\eta\right)^2}\right) \times$$

$$\times \left|\frac{d(a'_\nu b'_\omega - a'_\omega b'_\nu) + c'_\nu(ba'_\omega - ab'_\omega)}{d^2\sqrt{a^2+b^2}}\right|.$$

(3.12)

The results of calculations by the formula (3.12) are shown below in Fig. 11a (Appendix 9).

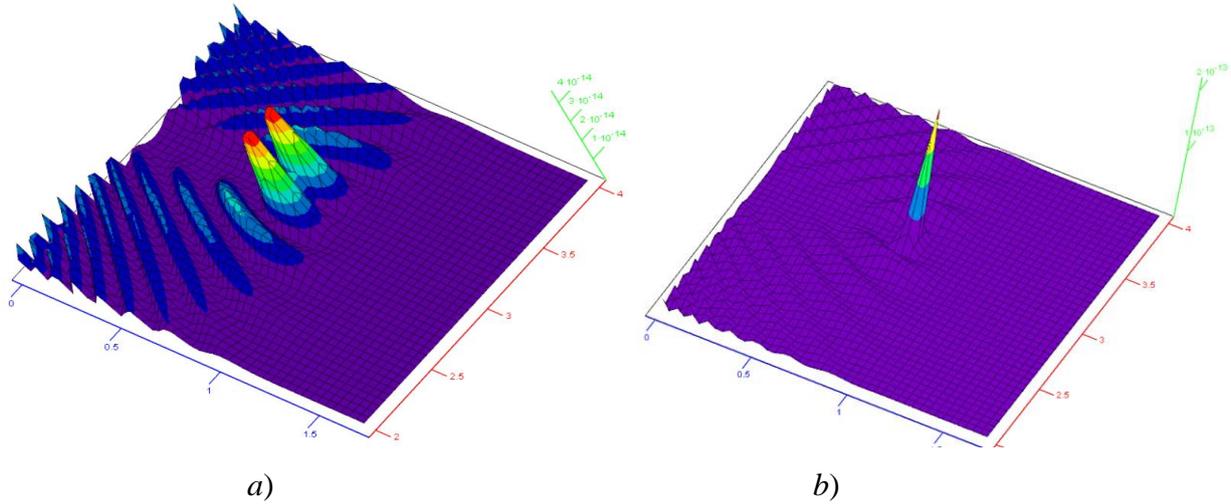

*a)*          *b)*

**Fig. 11a** Volumetric DESM on a multilayer non-isotropic crystal surface calculated by the formula (3.12) at $\vartheta = 45^0$, $\gamma = 0^0$, $l_1 = 10^{-11}$cm, $r_{cor5} = 4\times 10^{-9}$cm, *a)* $n_1 = 48$ and *b)* $n_1 = 47$

If each crystal layer has the same anisotropy, for example, of type (2.23a), then, taking into account (2.24a) and (2.28a), we obtain the following formula for calculating volumetric DESM for this case

$$D(v,\omega/\vartheta,\gamma) = \frac{4}{\pi^2 l_2} \left( \frac{a^2 b^2}{(a^2+b^2)^2} \right) \left( \frac{\cos^2(\pi n_1) - \cos(\pi n_1)\cos\left(\sqrt{\frac{a^2+b^2}{d^2}} l_2/\eta\right)}{\left(\pi n_1/l_2\right)^2 - \left(\sqrt{\frac{a^2+b^2}{d^2}} l_2/\eta\right)^2} - \frac{\cos\left(\pi n_1 + \sqrt{\frac{a^2+b^2}{d^2}} l_2/\eta\right) - 1}{\left(\pi n_1/l_2 + \sqrt{\frac{a^2+b^2}{d^2}}/\eta\right)^2} \right) \times$$

$$\times \left| \frac{d(a'_v b'_\omega - a'_\omega b'_v) + c'_v (ba'_\omega - ab'_\omega)}{d^2 \sqrt{a^2+b^2}} \right|.$$

(3.12a)

The results of calculations by the formula (3.12a) are shown in Fig. 11*b* (see Appendix 10).

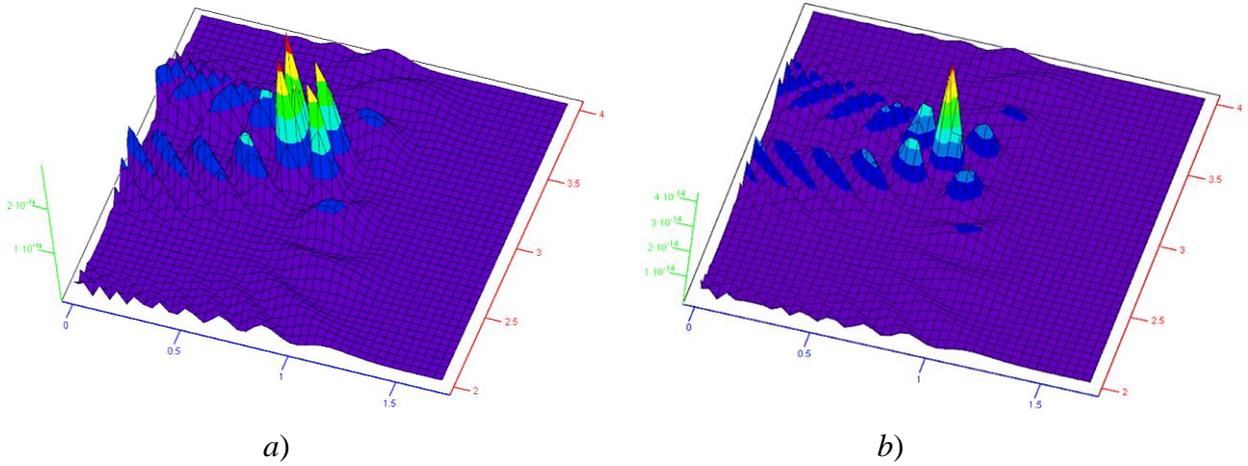

*a*)                                                                                   *b*)

**Fig. 11a** Volumetric DESM on a multilayer non-isotropic crystal surface calculated by the formula (3.12a) at $\vartheta = 45^0$, $\gamma = 0^0$, $l_1 = 10^{-11}$ cm, $r_{cor5} = 4 \times 10^{-9}$ cm, *a*) $n_1 = 42$ and *b*) $n_1 = 37$

Let's analyze the volumetric diagrams of elastic scattering of microparticles shown in Figures 11, 11*a* and 11*b*.

1] *Agreement with experiment*

A separate study should be devoted to comparing the calculations according to the formulas presented in this article with experimental data. But, already at this stage, it can be noted that volumetric DESMs calculated by the formula (3.9) (Figure 11) correspond to the results of experiments on the diffraction of particles and high-frequency electromagnetic waves on a crystal (see Figures 12, 12*a*).

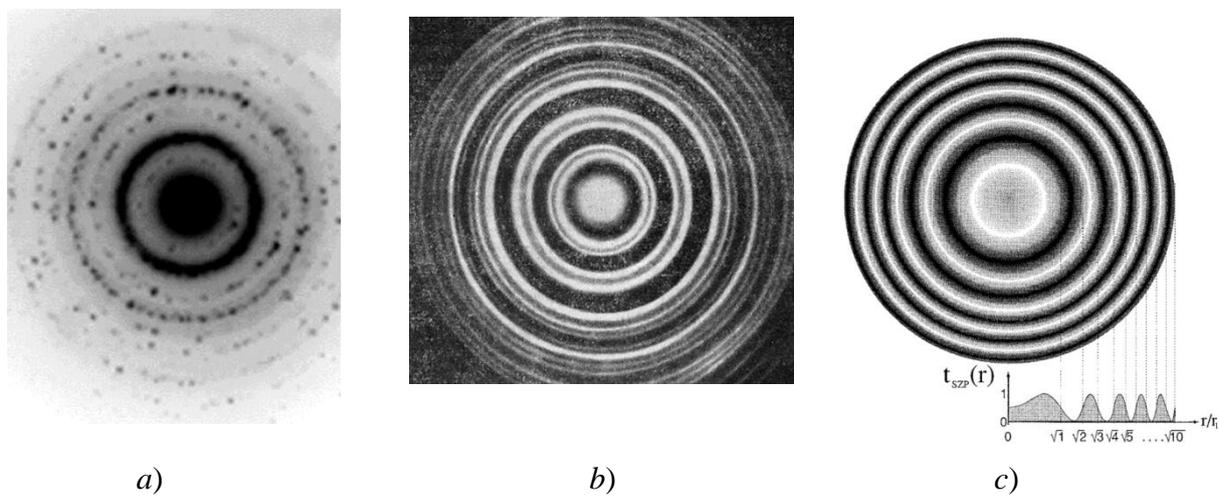

*a)* *b)* *c)*

**Fig. 12** *a*) Electron diffraction pattern of the Ti$_{50}$Ni$_{25}$Cu$_{25}$ alloy (http://dream-journal.org/issues/2018-6/2018-6_233.html); *b*) Electron diffraction on gold. The thickness of the gold plate was about 250Å = 2.5×10$^{-6}$cm. The size of the gold atom is approximately equal to 0,28 nm = 2.8×10$^{-8}$cm. Thus, in the gold plate there were approximately 100 layers (i.e., atomic planes); *b*) Illustration of an *X*-ray diffraction pattern obtained by diffraction of photons on a crystal. Photos and drawing are taken from the World Wide Web in the public domain

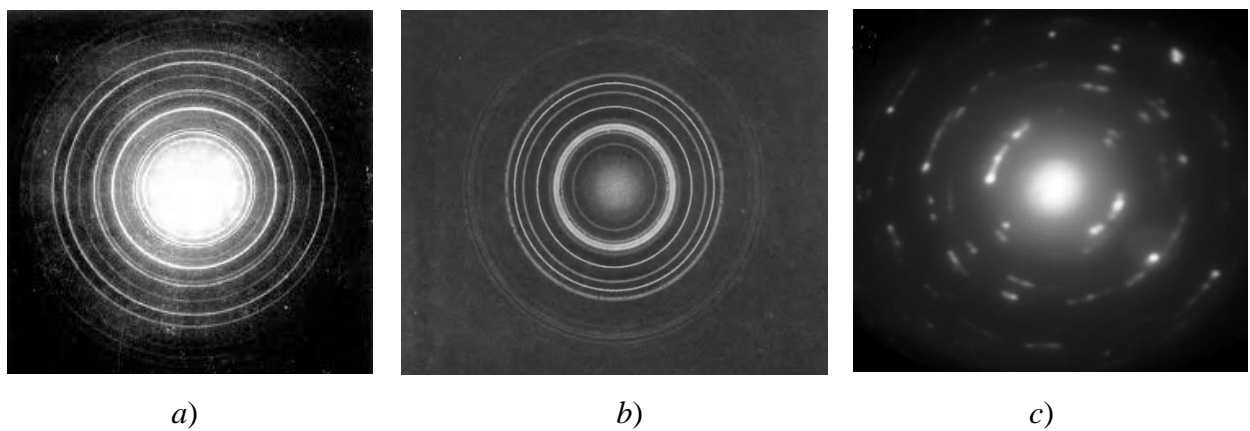

*a)* *b)* *c)*

Fig. 12a *a*) Electron diffraction pattern of the NaCl standard; *b*) Electron diffraction pattern of a polycrystal of hexagonal nickel hydride NiH$_2$ (http://ignorik.ru/docs/lekciya-13-eksperimentalenie-metodi-kristallofiziki.html); *c*) Electron diffraction on aluminum Al. https://www.researchgate.net/publication/295974108_Electron_Diffraction. Photos are taken from the World Wide Web in the public domain

Formula (3.9) has a significant advantage in that it allows a more subtle analysis of the process of scattering of microparticles on a crystal than methods based on the idea of the existence of de Broglie waves. A selection of the parameters $\vartheta$, $l_1$, $n_1$ and $r_{cor5}$ can lead to similarity with experimentally obtained electron diffraction patterns or *X*-ray diffraction patterns, and more detailed information on the structure of the crystal or other multilayer reflecting surface is disclosed.

2] *"Adjustment" of the scale parameter $\eta$*

To take into account the various features of the crystal lattice, a scale parameter (2.78)

$$\eta = \frac{l_1^2(\pi^2 n_1^2 - 6)}{6\pi^2 r_{cor5}}$$

can be "adjusted" to the results of experiments. For example, can change the values of numeric constants and / or enter functional dependencies on parameters $l_1$, $n_1$ and $r_{cor5}$:

$$\eta = \frac{l_1^2(\pi^2 n_1^2 - 3)}{12\pi^2 r_{cor5}} \quad \text{or} \quad \eta = \frac{l_1^2(\pi^2 n_1^2 - 127)}{16\pi^2 r_{cor5}} \quad \text{or} \quad \eta = \frac{l_1^2(4\pi^2 n_1^2 - 33)}{12\pi^2 r_{cor5}} \quad \text{and etc.} \quad (3.13)$$

$$\eta = \frac{l_1^2[\pi^2 \cos^2(\pi n_1/N_1) - 8]}{4\pi^2 r_{cor5}} \quad \text{or} \quad \eta = \frac{\ln(l_1^2)[\pi^2 tg^2(\pi n_1/N_1) - 13]}{7\pi^2 r_{cor5}} \quad \text{and etc.}$$

Perhaps such an "adjustment" $\eta$ will lead to greater similarity of the results of calculations by the formula (3.9) with real electron diffraction patterns or radiographs. At the same time, the "adjustment" of the scale parameter (2.78) can make it possible to evaluate additional features of the structure and / or defects of the crystal lattice.

When "adjusting" $\eta$, however, it is necessary to consider that equation (3.9) must satisfy the condition

$$\int_0^{\frac{\pi}{2}} \int_0^{2\pi} \rho(v, \omega/\vartheta, \gamma)\, dv\, d\omega = 1, \quad (3.14)$$

where the angle $v$ varies from 0 to $\pi/2$; the angle $\omega$ varies from 0 to $2\pi$.

3] *Even and odd number of crystal layers*

From the diagrams shown in Figure 11, it can be seen that if the even number of layers $n_1$ effectively involved in the reflection of microparticles, then a minimum (dip) is observed in the very center of the diagram; and if the number of reflecting layers is odd, then a maximum (peak) is observed in the very center of the diagram. The same effect is found in experiments (Figure13).

It should be noted, however, that for $n_1 = 4$ (i.e., for an even number of layers) in the center of the diagram, there is not a minimum, but a maximum (peak) (Figure 20).

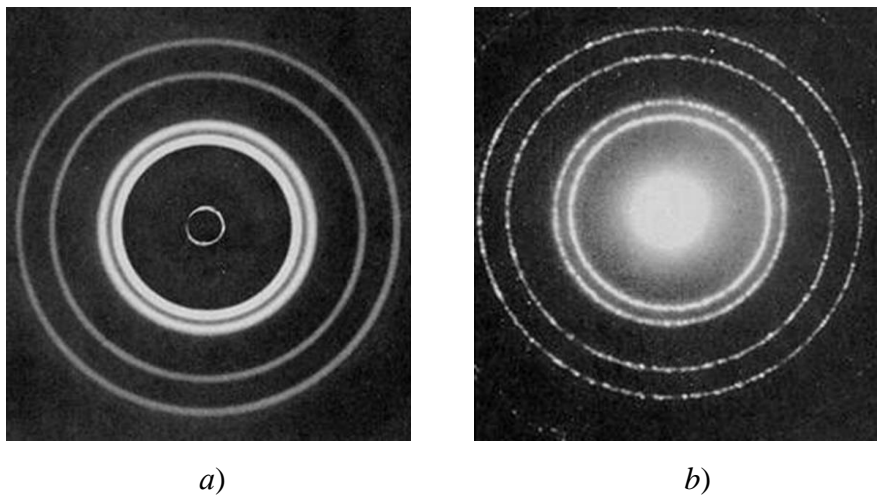

*a)*          *b)*

**Fig. 13** *a)* In a number of experiments on the diffraction of microparticles, a dark spot is observed in the center of the electron diffraction pattern or X-ray diffraction pattern. *b)* In a number of other similar experiments, a bright spot is observed in the center of the electron diffraction pattern or radiograph

4] *The falling velocity v of microparticles and the number of reflective layers $n_1$*

It should be expected that the number of layers $n_1$, which are penetrated by microparticles incident to the surface of the crystal, mainly depends on their energy $E$ [i.e. $n_1 = f(E)$]. More precisely, for incident fermions (in particular, electrons), the immersion depth in the thickness of the reflecting surface (i.e., the number of layers $n_1$) mainly depends on their speed (the momentum or kinetic energy), and for incident bosons (in particular, photons) from their frequency. More generally, it is possible to find the dependence

$$n_1 = f(E, l_1, r_{cor5}, \vartheta, \gamma). \tag{3.15}$$

The expression (3.15) can also take into account the effects of shading part of the deep sections of the reflecting surface at small angles $\vartheta$, etc.

The determination of the functional dependence (3.15) will make it possible to more accurately match the results of calculations by the formula (3.9) with experimental data on the diffraction of microparticles on periodic structures such as crystals and to obtain additional information about the structure of the reflecting surface.

In particular, we consider the DESM (3.9) as a function of the number of layers $n_1$ of the reflecting surface of the single crystal $D(n_1)$ with the six fixed parameters $\vartheta$, $\gamma$, v, $\omega$, $l_1$, $r_{cor5}$. The results of calculations by the formula (3.9) $D(n_1)$ in this case are shown in Fig. 14a (Append

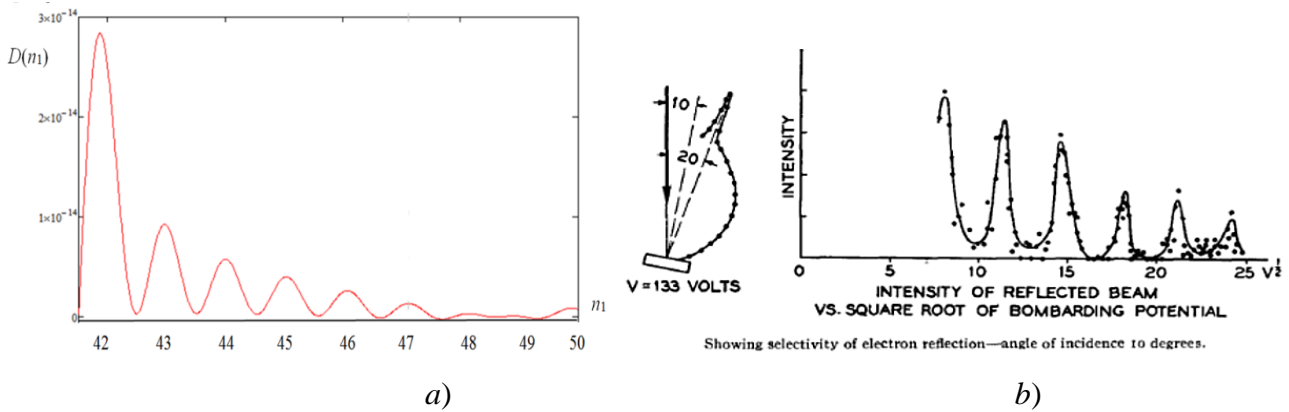

a)  b)

**Fig. 14** *a*) Dependence of the DESM (3.9) on the number of layers $n_1$ of the reflecting surface of the single crystal, which, in turn, depends on the velocity $v$ (more precisely, energy E) of the microparticles incident on this surface (3.15). The calculations were performed according to the formula (3.9) $D(n_1)$ as a function of the number $n_1$, which varies in the range from 40 to 50 layers, with the following constant parameters: $\vartheta = 45^0$, $\gamma = 0^0$, $v = 45^0$, $\omega = 0^0$, $l_1 = 10^{-11}$cm, $r_{cor5} = 9\times10^{-9}$cm;  *b*) The intensity of an electron beam *I* scattered on a nickel single crystal at a constant reflection angle, depending on the square root of the voltage *U*, accelerating particles in an electron gun (electron generator). This experimental dependence was first obtained in 1927 by Clinton Davisson and Lester Germer [1]

Considering that the number of crystal layers penetrated by incident microparticles depends on their speed $n_1 = f(v)$, these calculations using the formula (3.9) $D(n_1)$ are in good agreement with the results of K. Davisson and L. Germer's experiment (1927) on electron diffraction on a nickel crystal [1] (Figure 14*b*).

Formula (3.9) $D(n_1)$ allows one to perform calculations in a much wider range of $n_1$ values (Figure 15). At the same time, Figure 15*a* shows that the in range $n_1$ from 0 to 40 layers $D(n_1)$ can take negative values. Since formula (3.9) $D(n_1)$ is an OPDF, then at first glance this looks like an absurd result.

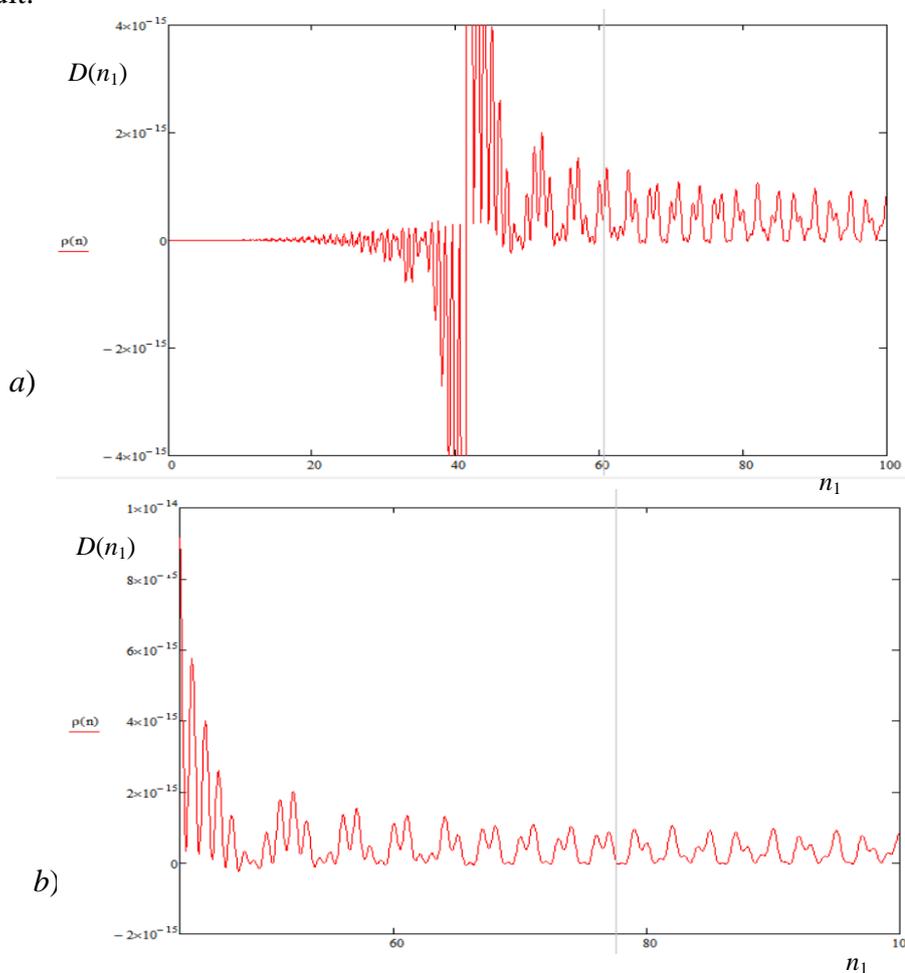

*a)*

*b)*

**Fig. 15** Results of calculations by the formula (3.9) $D(n_1)$ as a function of the number of layers $n_1$, varying in the range from *a)* 0 to 100 layers; *b)* 40 to 100 layers, with the following constant parameters $\vartheta = 45^0$, $\gamma = 0^0$, $v = 45^0$, $\omega = 0^0$, $l_1 = 10^{-11}$cm, $r_{cor5} = 9 \times 10^{-9}$cm. Calculations performed using MathCad software

Negative values in the calculations by formula (3.9) can be "eliminated" by assuming that in the case under consideration (that is, with $l_1 = 10^{-11}$ cm и $r_{cor5} = 9 \times 10^{-9}$ cm), the depth of the reflecting layer cannot be less than $l_2 = l_1 n_1 = 40 \times 10^{-11} = 4 \times 10^{-10}$ cm.

In another case, when $l_1 = 10^{-11}$ cm and $r_{cor5} = 2 \times 10^{-11}$ cm - corresponds to the size of an atom that effectively reflects electrons or high-frequency photons, a calculation by the formula (3.9) $D(n_1)$ leads to the result shown in Figure 16. In this case, the prohibition applies only to $l_2$ in the 3 through 4 first layers.

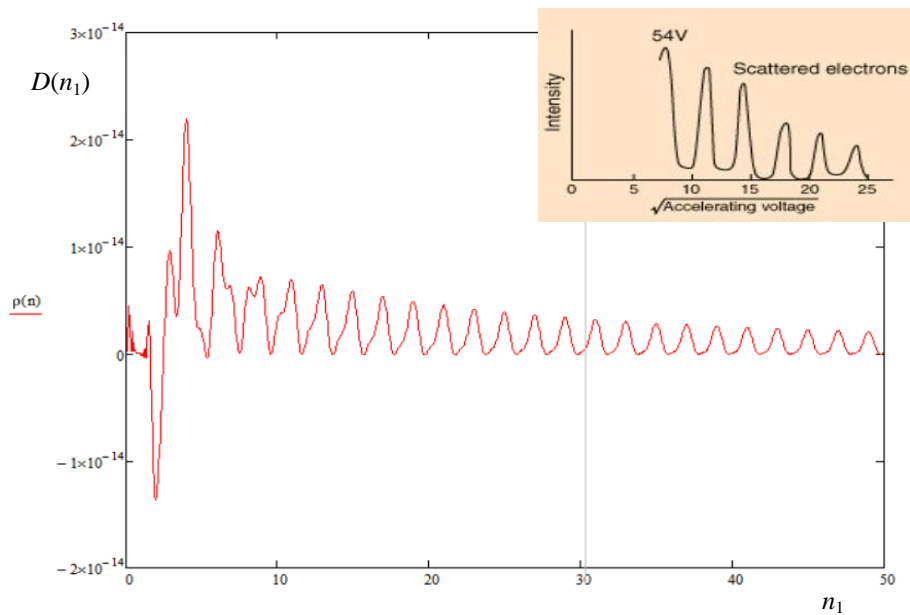

**Fig. 16** The result of the calculation by the formula (3.9) $D(n_1)$ for the following unchanged parameters $\vartheta = 45^0$, $\gamma = 0^0$, $v = 45^0$, $\omega = 0^0$, $l_1 = 10^{-11}$ cm, $r_{cor5} = 2 \times 10^{-11}$ cm, and the experimental dependence obtained by K. Davisson and L. Germer in the study of electron diffraction on a nickel crystal [1]

On the other hand, as will be shown below, the negative results of calculations by the formula (3.9) $D(n_1)$ can mean that, when scattering microparticles on the thin films (i.e., for $n_1 < 12$), part of the microparticles pass through the atomic lattice.

In this article, the problem $D(n_1) < 0$ has no final solution. A separate theoretical and experimental study should be devoted to this issue.

*5] Scattering of microparticles on a single crystal layer*

When scattering microparticles on one layer of the crystal (i.e., for $n_1 = 1$), the calculation by the formula (3.9) leads to the result shown in Figure 17*a,b*.

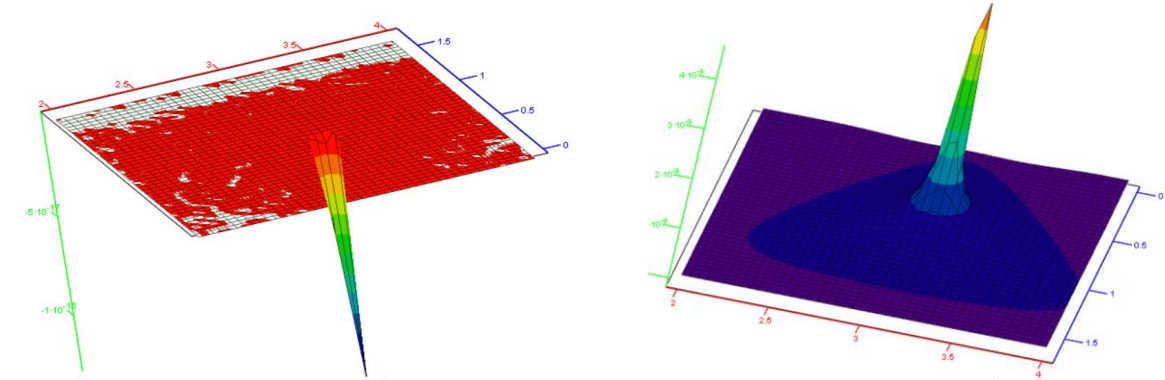

a) $n_1 = 1$, $\vartheta = 45^0$, $\gamma = 0^0$,           b) $n_1 = 1$, $\vartheta = 45^0$, $\gamma = 0^0$,
$l_1 = 10^{-11}$cm, $r_{cor5} = 6 \times 10^{-9}$cm      $l_1 = 10^{-8}$ cm, $r_{cor5} = 6 \times 10^{-9}$cm

**Fig. 17** Diagrams of elastic scattering of microparticles (DESM) on one layer of the crystal ($n_1 = 1$), calculated by the formula (3.9) for various $l_1$

If the thickness of the first layer is $l_1 = 10^{-11}$cm, then the calculation result by the formula (3.9) is negative (Figure 17*a*). This can be explained by the fact that microparticles do not reflect from this layer, but pass through it. If the first layer is thicker, for example, $l_1 = 10^{-8}$ cm, then the reflection from such a layer (Figure 17*b*) is similar to the reflection from the top layer of an uneven surface with other statistics of the height of irregularities (Figures 7 through 10).

An interesting calculation results using the formula (3.9) is observed for $n_1 = 1$ and $l_1 = 4 \times 10^{-10}$cm (Figure 18). This case can be interpreted as a prediction that part of the microparticles will reflect from one layer of the crystal, and the other part of the microparticles will pass through it.

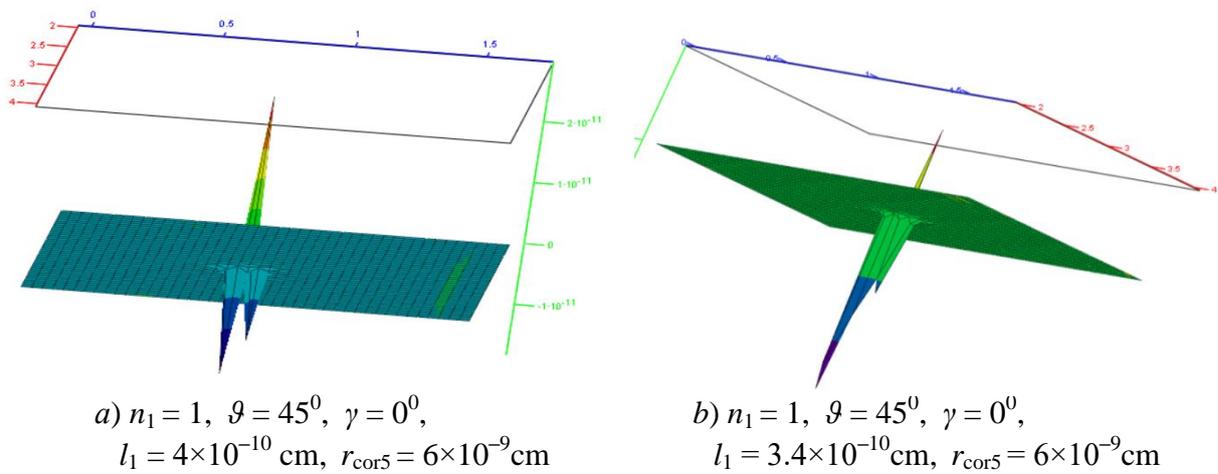

a) $n_1 = 1$, $\vartheta = 45^0$, $\gamma = 0^0$,
$l_1 = 4\times10^{-10}$ cm, $r_{cor5} = 6\times10^{-9}$ cm

b) $n_1 = 1$, $\vartheta = 45^0$, $\gamma = 0^0$,
$l_1 = 3.4\times10^{-10}$ cm, $r_{cor5} = 6\times10^{-9}$ cm

*6] Scattering of microparticles on two, three and four layers of the crystal*

Diagrams of elastic scattering of microparticles on two, three and four layers of the crystal, calculated by the formula (9.3), are shown in Figures 19 and 20.

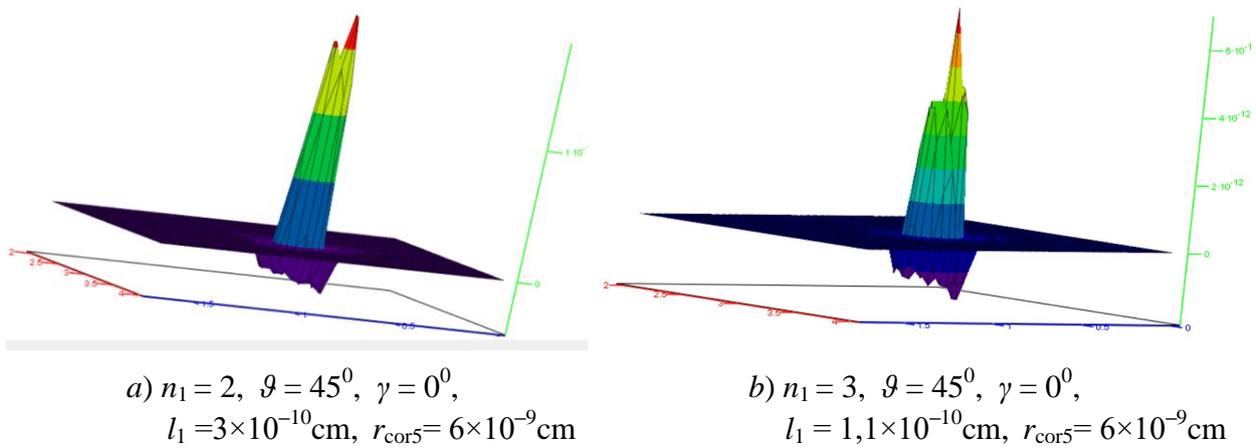

a) $n_1 = 2$, $\vartheta = 45^0$, $\gamma = 0^0$,
$l_1 = 3\times10^{-10}$ cm, $r_{cor5} = 6\times10^{-9}$ cm

b) $n_1 = 3$, $\vartheta = 45^0$, $\gamma = 0^0$,
$l_1 = 1,1\times10^{-10}$ cm, $r_{cor5} = 6\times10^{-9}$ cm

**Fig. 19** DESM on two (*a*) and three (*b*) layers of the crystal, calculated by the formula (3.9)

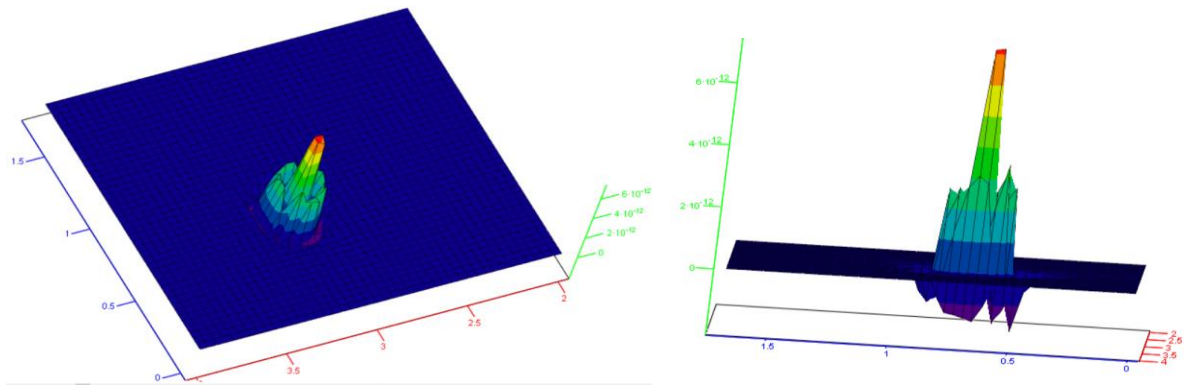

**Fig. 20** Two angles of the DESM on four layers of the crystal, calculated by the formula (3.9) for $n_1 = 4$, $\vartheta = 45^0$, $\gamma = 0^0$, $l_1 = 1{,}2 \times 10^{-10}$cm, $r_{cor5} = 9 \times 10^{-9}$cm

*7] The fifth parameter γ*
As shown above, by selecting four parameters:

H'   V   H   I
$\vartheta$,   $l_1$,   $n_1$,   $r_{cor5}$

it is possible to achieve that the calculations by the formula (3.9) correspond to different diffraction patterns of microparticles on a multilayer statistically uneven surface of the crystal.

The fifth parameter (quintessence from lat. *quīnta essentia* "fifth essence") is the angle $\gamma$ (see Figures 3 and 5), in all the previously considered cases it remained equal to zero ($\gamma = 0^o$).

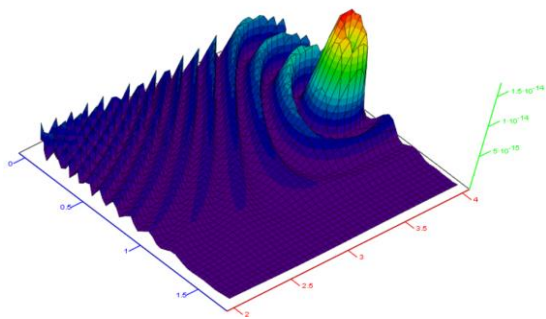
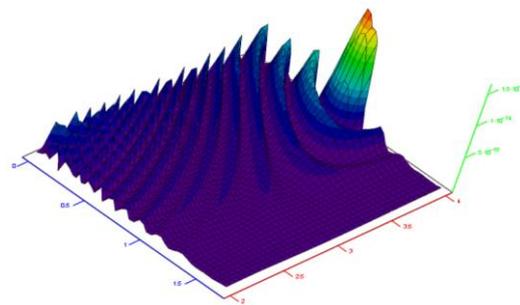

a) $n_1 = 66$, $\vartheta = 45^0$, $\gamma = 35^0$, $l_1 = 10^{-11}$см, $r_{cor5} = 6 \times 10^{-9}$см

b) $n_1 = 66$, $\vartheta = 45^0$, $\gamma = 55^0$, $l_1 = 10^{-11}$см, $r_{cor5} = 6 \times 10^{-9}$см

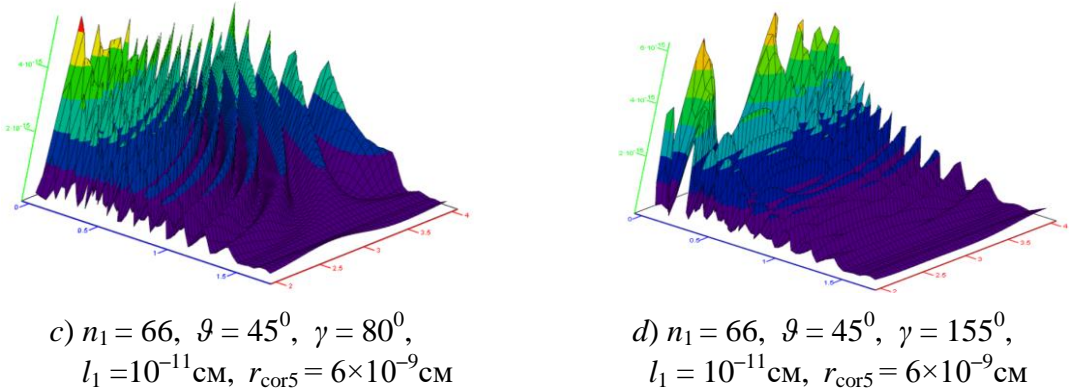

c) $n_1 = 66$, $\vartheta = 45^0$, $\gamma = 80^0$, $l_1 = 10^{-11}$ см, $r_{cor5} = 6\times 10^{-9}$ см

d) $n_1 = 66$, $\vartheta = 45^0$, $\gamma = 155^0$, $l_1 = 10^{-11}$ см, $r_{cor5} = 6\times 10^{-9}$ см

**Fig. 21** DESM on a crystal, calculated by the formula (3.9), for the identical $\vartheta$, $n_1$, $l_1$, $r_{cor5}$ and different angles $\gamma$

When deriving the formula (3.9), it was taken into account that all azimuthal cross-sections in different directions of a homogeneous and isotropic uneven surface of the crystal are the same. Therefore, it was expected that when the azimuthal angle $\gamma$ changes, the scattering diagram should remain unchanged, and only its azimuthal direction should change. From the diagrams shown in Figure 21*a,b*, it can be seen that for small angles $\gamma$ equal to $35^0$ and $55^0$, only the azimuthal direction of the whole diagram shifts. But with a further increase in the angle $\gamma$, the scattering diagram changes significantly with the remaining four parameters $\vartheta$, $l_1$, $n_1$, $r_{cor5}$ unchanged (Figure 21 *c,d*).

At this stage of the study, it is difficult to establish whether this change is a drawback of the formula (3.9), or is it a reflection of reality that can be experimentally confirmed.

It can be assumed that the DESM depends on the angle α between the projection of the azimuthal direction of motion of the incident microparticles on the XOY-plane and the direction of the rows of atoms in the crystal lattice (Figure 22).

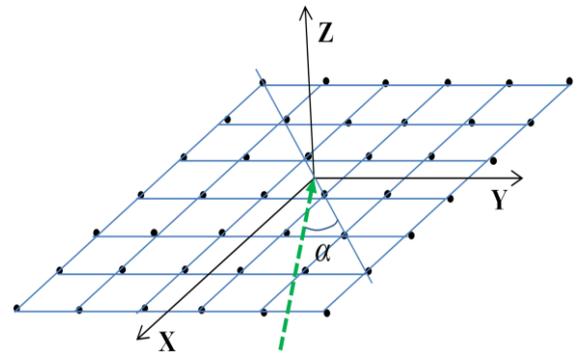

**Fig. 22** Angle *α* between the projection of the azimuthal direction of motion of incident microparticles on the XOY-plane and the direction of the rows of atoms in the crystal lattice

From Figure 22 can be seen that rotation of the plane of incidence of the microparticles at the angle

$\alpha$ is accompanied by the effect of increasing the distance between the atoms of the crystal lattice, which are effectively involved in their scattering. This effect can be taken into account by increasing the correlation radius of the heights of the surface irregularities $r_{cor5}$. The scattering diagrams at $\gamma = 75^0$ and enlarged in comparison with the previous case of $r_{cor5}$ and $l_1$ are shown in Figure 23.

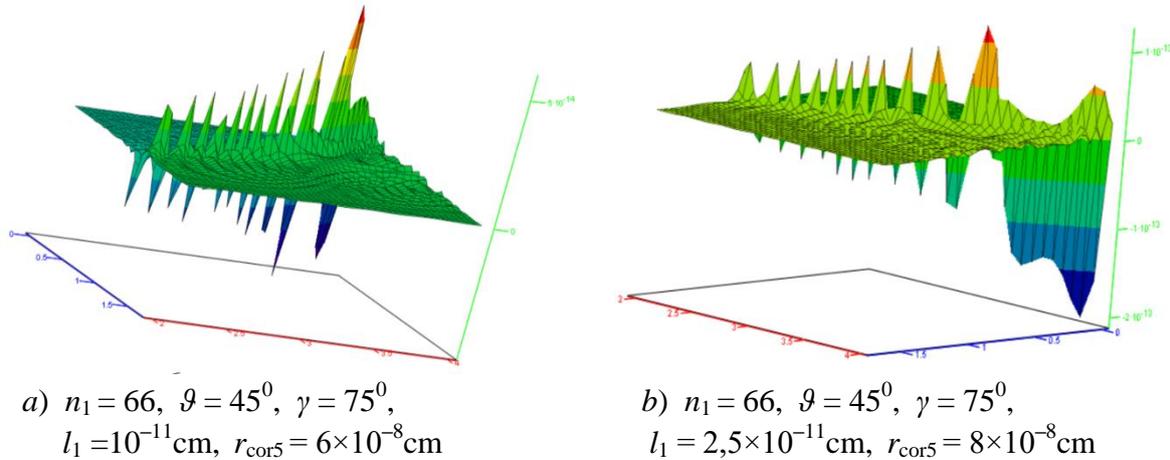

*a)* $n_1 = 66$, $\vartheta = 45^0$, $\gamma = 75^0$, $l_1 = 10^{-11}$cm, $r_{cor5} = 6 \times 10^{-8}$cm

*b)* $n_1 = 66$, $\vartheta = 45^0$, $\gamma = 75^0$, $l_1 = 2,5 \times 10^{-11}$cm, $r_{cor5} = 8 \times 10^{-8}$cm

**Fig. 23** DESM calculated by the formula (3.9), at $\gamma = 75^0$ and increased $r_{cor5}$ and $l_1$

These calculation results by the formula (3.9) are subject to experimental verification. If the distortions of the DESM due to a change in the angle γ are not experimentally confirmed, then this disadvantage can be compensated for by a change in the orientation of the reference frame. In many cases, the coordinate axis from which the angle *γ* is measured can be initially combined with the azimuthal direction of motion of the microparticles incident on the crystal surface. That is, in a number of experiments, taking advantage of the arbitrariness in choosing a reference frame, it is possible from the very beginning to achieve that $\gamma = 0^0$.

8] *Diffraction of microparticles on thin films*

The DESM calculation procedure presented in §2.1 and §2.5 was developed on the basis that microparticles, after a collision with a solid surface, are reflected from it according to the laws of geometric optics, and do not pass through this body. But it turned out that the formula (3.9) makes it possible to calculate the scattering diagram when microparticles pass through thin films.

Figure 24 shows the scattering diagrams of microparticles on thin films consisting of 14 and 15 layers of the crystal lattice.

Diffraction maxima are obtained when microparticles fall on thin films at angles $\vartheta$ from $25^0$ to $65^0$. In this case, some of the microparticles are reflected from the uneven layers (i.e., atoms) of the thin film, and the other part passes through them.

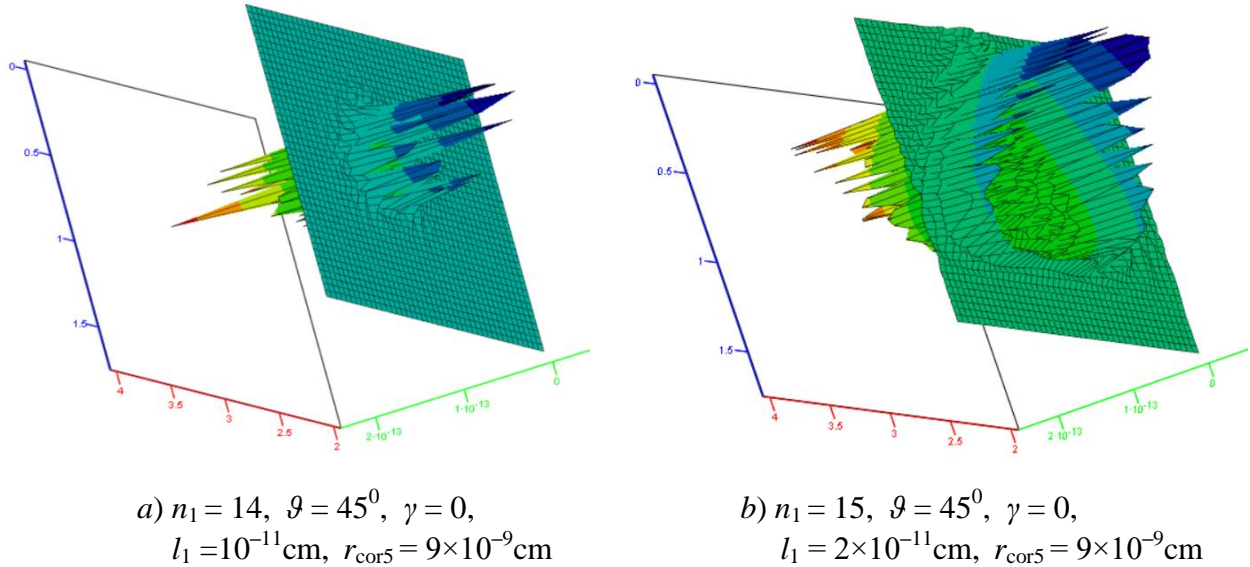

a) $n_1 = 14$, $\vartheta = 45^0$, $\gamma = 0$, $l_1 = 10^{-11}$cm, $r_{cor5} = 9 \times 10^{-9}$cm

b) $n_1 = 15$, $\vartheta = 45^0$, $\gamma = 0$, $l_1 = 2 \times 10^{-11}$cm, $r_{cor5} = 9 \times 10^{-9}$cm

**Fig. 24** Diffraction maxima of microparticles passing through thin films calculated by the formula (3.9)

When microparticles fall vertically on the surface (i.e., at $\vartheta = 90^0$), calculations using the formula (3.9) lead to absurd results. In other words, the method of calculating DESM proposed in this article does not apply to this case. It should be noted that diffraction maxima are obtained when microparticles fall on thin films at angles $\vartheta$ from $25^0$ to $65^0$. In this case, some of the microparticles are reflected from the uneven layers (i.e., atoms) of the thin film, and the other part passes through them.

9] *Overall remarks*

Summarizing this section, we note that the formula (3.9) [or in another form (3.10)] opens up wide opportunities for studying the properties of solid materials by analyzing the results of scattering of microparticles on them.

By selecting five parameters $\vartheta$, $l_1$, $n_1$, $r_{cor5}$ and $\gamma$, which are associated with some properties of the atomic or molecular structure of a solid, one can achieve a similarity of the scattering diagram calculated by the formula (3.9) with an electron diffraction or X-ray, and thereby obtain information about the structure of this body.

In general, the formula (9.3) with the five parameters $\vartheta$, $l_1$, $n_1$, $r_{cor5}$ and $\gamma$ generates an infinite set of two-dimensional surfaces in which individual forms can exist that reflect the outlines or essence of processes in the surrounding reality. However, all these surfaces have the following common property. Since the formula (3.9) is the one-dimensional probability density function (OPDF) $D(v,\omega/\vartheta,\gamma) = \rho(v,\omega/\vartheta,\gamma)|G_{v\omega}|$, the total area of all these surfaces is equal to one (3.14).

The formula (3.9) is suitable for describing elastic diffraction not only of elementary particles, atoms, and photons, but also for scattering macroscopic elastic bodies (such as soccer balls or tennis balls) on large multi-layer periodic structures. Let, for example, a three-dimensional grid with the edge length of one cubic cell of 3400 cm = 34 m be assembled from metal pipes with a diameter of 30 through 50 cm, whereby metal balls with a diameter of 50 to 80 cm are placed in the nodes of this grid (see Figure 25).

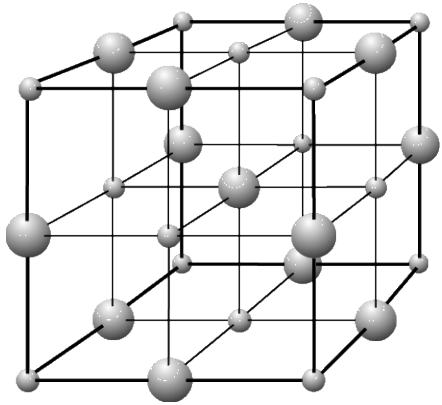

**Fig. 25** Cubic lattice consisting of metal pipes and balls of different diameters

If a stream of soccer balls with a diameter of 22.3 cm is directed at an angle $\vartheta = 45^0$ at such a cubic lattice, then their scattering is also described by the formula (3.9). Indeed, if instead of $l_1 = 10^{-11}$ cm, $r_{cor5} = 6 \times 10^{-9}$ cm and $n_1 = 66$, substitute $l_1 = 50$ см, $r_{cor5} = 3400$ cm and $n_1 = 18$ in the scale parameter $\eta$ (3.11), then the diagram of elastic scattering of soccer balls on such a cubic lattice, calculated by the formula (3.9), will be approximately the same as shown in Figure 11*a*.

If the case of diffraction of soccer balls is confirmed experimentally, then we can argue that the formula (3.9) turned out to be universal with respect to the different scales of the events studied, and the phenomena of the microworld are indistinguishable from the phenomena of the macrocosm (under similar conditions).

It is possible to pose the inverse problem of simulating processes occurring in the microworld using similar processes of the macrocosm. This will allow a more detailed understanding of the essence of microscopic phenomena.

**4 Summary**

The following results are obtained in this article.

**4.1 Method for calculating elastic scattering diagrams of microparticles**

In §§2.1 through 2.5, a method has been developed for calculating volumetric DESM on statistically uneven surfaces with various statistics of the height of irregularities. This method is applicable for describing the scattering of elastic particles and waves (photons and phonons), under the conditions of the Kirchhoff approximation:

- irregularities of the reflecting surface are statistically uniform, smooth and large-scale in comparison with the sizes of microparticles (their radius or wavelength);

- reflection of microparticles from all local sections of an uneven surface occurs according to the laws of geometric optics. For brevity, such a reflection of microparticles in the article is called "elastic";

- the section of the uneven reflecting surface is at a great distance from the generator and detector of the microparticles (see Figure 2).

In this work, attention is focused on the scattering of elementary particles (in particular, electrons and high-frequency photons). However, the article suggests that the proposed method is suitable for describing elastic scattering of large-scale bodies as well (for example, soccer or tennis balls), if the above conditions are met. In other words, it is assumed that there are no funda-mental differences between the diffraction of particles of the microworld and compact elastic bodies of the macroworld under similar conditions.

**4.2 OPDF derivative of a stationary random process**

Based on the procedure (2.29) through (2.32) given in [34, 35], in this article, we obtained:

1) an OPDF of the derivative of a Gaussian stationary random process (SRP) (2.40);

2) an OPDF derivative of the SRP with a uniform distribution of the heights of irregularities (2.51);

3) an OPDF derivative of the SRP with the Laplace distribution of the height of irregularities (2.65);

4) an OPDF derivative of the SRP with the distribution of the height of irregularities according to the Cauchy law (2.67);

5) an OPDF derivative of the SRP with the distribution of the heights of the irregularities according to the multilayer sinusoidal law (2.76).

The obtained OPDF $\rho[\xi'(r)]$ of derivatives of various SRP can be of interest for many branches of statistical physics. For example, since the momentum of a particle moving in the direction of the x-axis is related to the derivative of its coordinate by the ratio

$$p_x = mv_x = mdx/dt = mx',$$

the procedure (2.29) through (2.32) essentially means a transition from the coordinate a real representation of the statistical system, to its impulsive representation, with all the many consequences arising from this.

**4.3 The diagrams of elastic scattering of microparticles on a single layer uneven surface**

Based on the method described in §§ 2.1 through 2.5 and the OPDF of derivatives of stationary random processes obtained in §2.6, the following formulas are derived for calculating the elastic scattering diagrams of microparticles on single-layer, large-scale uneven surfaces:

1) a formula for calculating DESM $D(v,\omega/\vartheta,\gamma)$ on a single-layer surface with Gaussian distribution of the heights of irregularities (3.3);

2) a formula for calculating DESM $D(v,\omega/\vartheta,\gamma)$ on a single-layer surface with a uniform distribution of the heights of irregularities (3.4);

3) a formula for calculating DESM $D(v,\omega/\vartheta,\gamma)$ on a single-layer surface with a Laplace distribution of the heights of irregularities (3.5);

4) a formula for calculating DESM $D(v,\omega/\vartheta,\gamma)$ on a single-layer surface with the distribution of the heights of irregularities according to Cauchy's law (3.6).

Due to limitations imposed on the length of the paper, there is no detailed comparison of volumetric DESM calculated by the formulas (3.3) through (3.6) with experimental data. However, we note that in some cases the obtained DESMs are in good agreement with experiments (under the conditions of the Kirchhoff approximation) described in the extensive literature on the scattering of waves and particles on statistically uneven surfaces [17 through 27]. For example, Figure 26 compares the DESM calculated by formula (3.5) with the experimentally obtained diagram of neutron scattering on a single crystal $CsHSeO_4$ [36].

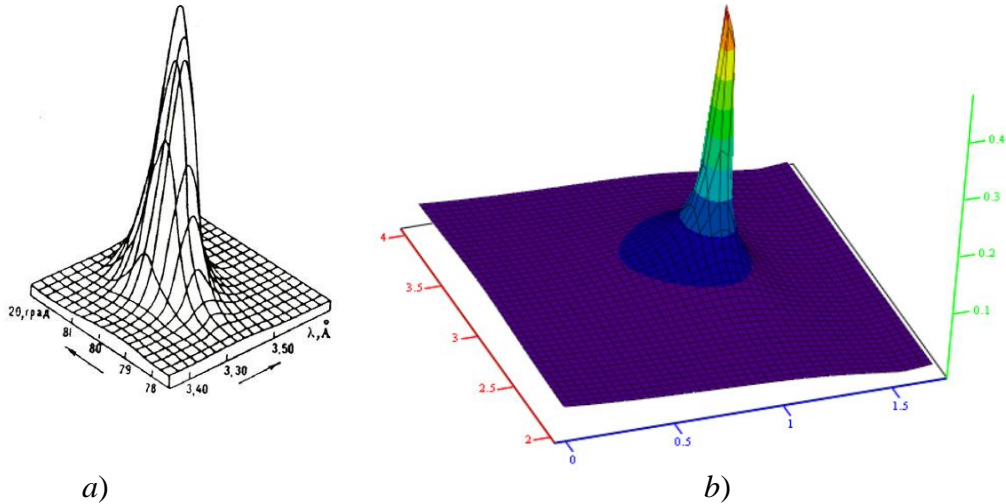

*a)* *b)*

**Fig. 26** *a*) The diffraction maximum of the neutron intensity reflected from the single crystal $CsHSeO_4$ [36]; *b*) Volumetric DESM, calculated by formula (3.5) for the case of the Laplace distribution of the heights of irregularities of reflecting surface, for $\vartheta = 60^0$, $\gamma = 0^0$, $\mu_L = 7$, $r_{cor3} = 5$

**4.4 The diagrams of elastic scattering of microparticles on a multilayer uneven surface**

On the basis of the method described in §§ 2.1 through 2.6, and the OPDF of the derivative of a multilayer sinusoidal stationary random process (2.76), in this paper we obtained the formula (3.9) for calculating the DESM on large-scale (compared to microparticles) irregularities of the multilayer surface of crystal.

By selecting the five parameters $\vartheta$, $l_1$, $n_1$, $r_{cor5}$ and $\gamma$ included in the equation (3.9), it is possible to achieve similarity of the scattering diagram of microparticles on the multilayer surface of the crystal calculated using this formula with experimentally obtained electron diffraction patterns (Figures 23, 24) or radiographs.

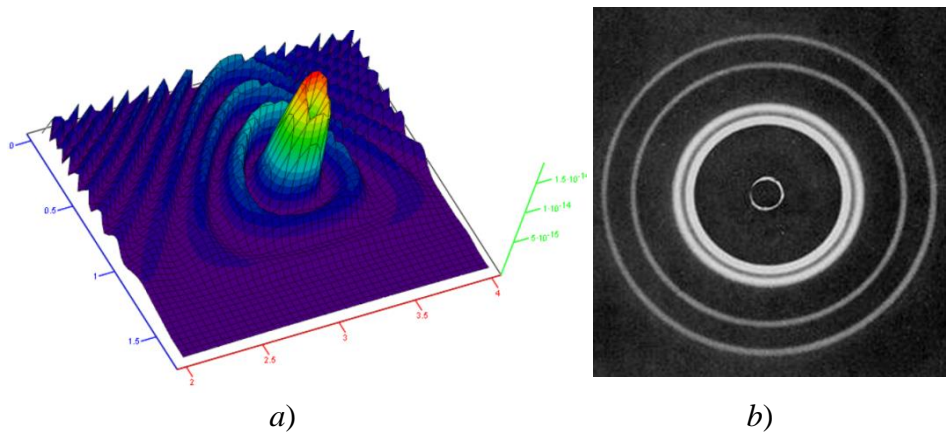

*a)*          *b)*

**Fig. 27** *a*) The volumetric diagram of the elastic scattering of microparticles on the multilayer surface of a crystal, calculated by the formula (3.9), for $\vartheta = 45^0$, $\gamma = 0^0$, $n_1 = 64$, $l_1 = 10^{-11}$cm, $r_{cor5} = 6 \times 10^{-9}$cm; *b*) Experimentally obtained electron diffraction pattern with a dark spot in the middle. Photo taken from a source that is freely available on the Internet

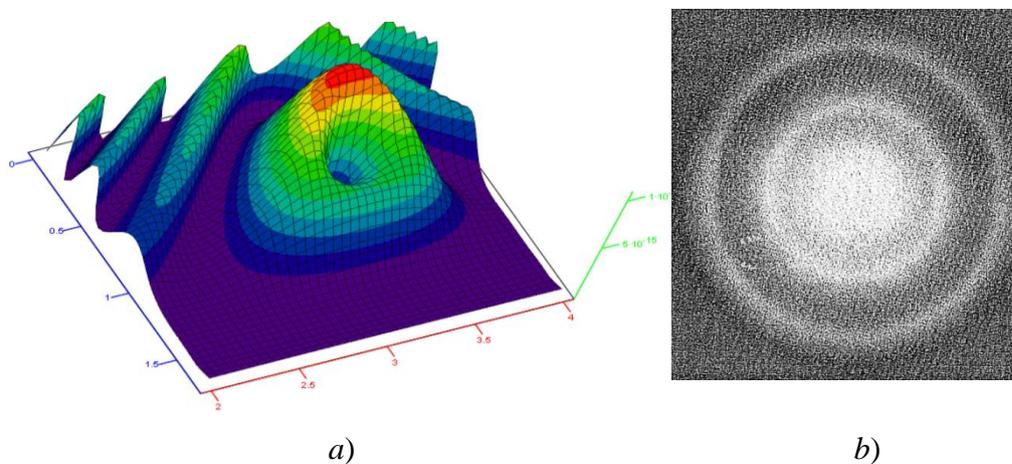

*a)*          *b)*

**Fig. 28** *a*) The volumetric diagram of the elastic scattering of microparticles on the multilayer surface of a crystal, calculated by the formula (3.9), for $\vartheta = 45^0$, $\gamma = 0^0$, $n_1 = 46$  $l_1 = 10^{-11}$cm, $r_{cor5} = 1.4 \times 10^{-9}$cm; *b*) Experimentally obtained electron diffraction pattern with a dark spot in the middle. Photo taken from a source that is freely available on the Internet.

Once again, we note that these results were obtained without using the idea of Louis de Broglie on the wave properties of elementary particles.

## 5 Conclusions

The article presents formulas for calculating the volumetric diagrams of elastic scattering of microparticles (DESM) (fermions and bosons) on uneven single-layer and multilayer surfaces with different statistics of the height of irregularities, when the conditions of the Kirchhoff approximation are met. At the same time, the one-dimensional probability density functions (OPDF) of the derivative of various stationary random processes are obtained, which can be used in a number of other problems of statistical physics.

In addition to solving the above practical problems, this article is aimed at introducing rational clarity into the conceptual problem associated with discussing the idea of the possible "existence" of de Broglie waves. The laws of geometric optics and the probabilistic methods of statistical physics applied here, according to the author, have allowed an explanation of the diffraction of elementary particles and atoms by crystals without using this hypothesis of Louis de Broglie about matter-waves. Moreover, this paper suggests that the phenomenon of particle diffraction on solid periodic structures can occur not only in the microcosm, but also in the macrocosm under similar conditions.

## 6 Acknowledgements

I thank my mentors, Dr. A.A. Kuznetsov and Dr. A.I. Kozlov for the formulation and discussion the problems outlined in this article. In performing the calculations, invaluable assistance was provided by Ph.D. S. V. Kostin. During the preparation of the manuscript, valuable remarks were made by D. Reed, Academician of the Russian Academy of Natural Sciences G.I. Shipov, Ph.D. V.A. Lukyanov and Ph.D. E. A. Gubarev, Ph.D. T.S. Levy.

Appendix 1

# The reflection of a plane electromagnetic wave
# from a square surface area

Let the length of a plane monochromatic electromagnetic wave $\lambda$ be much less than the characteristic dimensions of the surface irregularities of a solid or liquid substance conducting electric current (i.e., $\lambda \ll r_{cor}$, where $r_{cor}$ is the autocorrelation radius of the heights of the bumps in the reflecting surface). In this case, the uneven surface can be divided into many flat square sections (facets). Consider the reflection of the rays of the electromagnetic wave from each facet separately (Figure A.1.1 *a,b*).

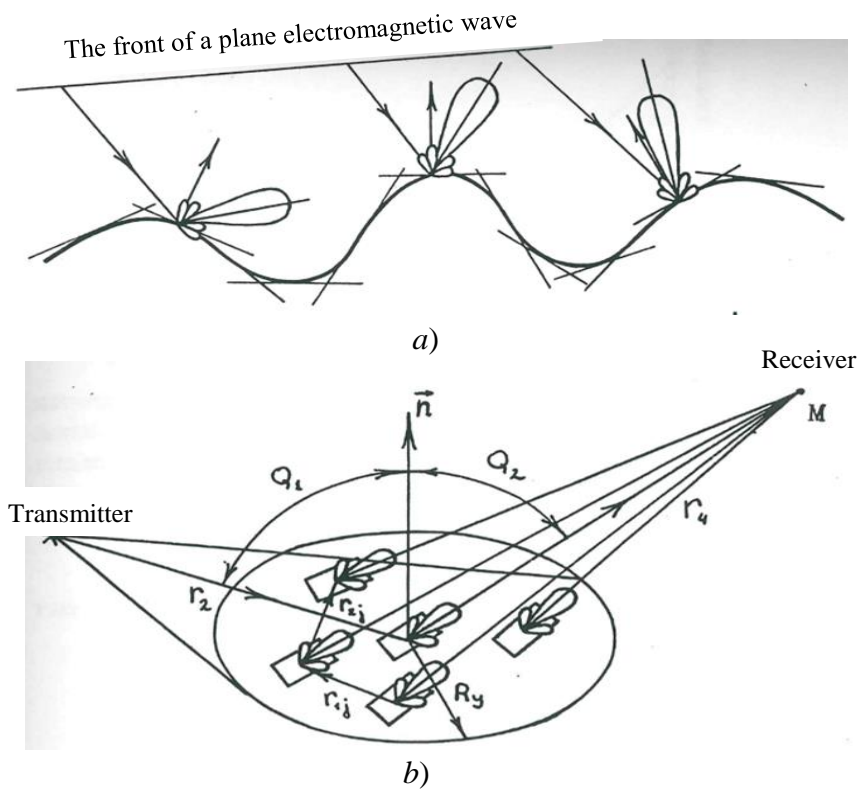

Fig. A.1.1 Scattering of an electromagnetic wave on a surface approximated by smooth square sections (facets). *a)* The maximum of the main lobe of the scattering diagram of each facet is directed according to the laws of geometric optics: lies in the plane of incidence and the angle of reflection is equal to the angle of incidence; *b)* Only the main lobes of the scattering diagrams whose facets are oriented accordingly are directed towards the receiver antenna

The beam of an electromagnetic wave here refers to a cylinder whose axis connects the source of the electromagnetic wave to the center of the reflecting facet, and the diameter of the base of this cylinder approximately coincides with the size of one of the sides of the $b_n$ square facet.

We define the scattering diagram of a flat monochromatic electromagnetic wave (EMW) on a single facet that perfectly conducts an electric current. Let's assume that the radiation point (emitter, Fig. A.1.1b) and the observation point (receiver) are located at a great distance from the facet (i.e. $b_n \ll r_2$ and $b_n \ll r_4$), so that the EMW rays incident on the facet and reflected from the facet can be considered almost parallel. In this case, the signal sent from any point on the square facet to the receiver antenna has the form

$$E_i(x, y) = \frac{E_m}{r_1} \exp\left\{i\left(\omega_1 t + \frac{2\pi}{\lambda}\left[\begin{array}{l} x(\cos\nu \sin\omega + \cos\vartheta \sin\gamma) + \\ + y(\cos\nu \cos\omega + \cos\vartheta \cos\gamma) - r_2 \end{array}\right]\right)\right\}, \quad (A.1.1)$$

where $x$ and $y$ determine the coordinates of each point on the square facet;

$E_m$ is the amplitude of the monochromatic electromagnetic field near the emitter;
$r_1$ is the distance from the source of EMW to the center of the facet (Figure A.1.1b);
$r_2$ is the distance from the center of the facet to the antenna of the receiver (Figure A.1.1b);
$\omega_1$ is the oscillation frequency of a monochromatic electromagnetic wave;
$\vartheta, \gamma$ are angles that specify the direction of the EMW ray incident on the facet (Figure A.1.2);
$\nu, \omega$ are the angles that specify the direction of the EMW ray reflected from the facet.

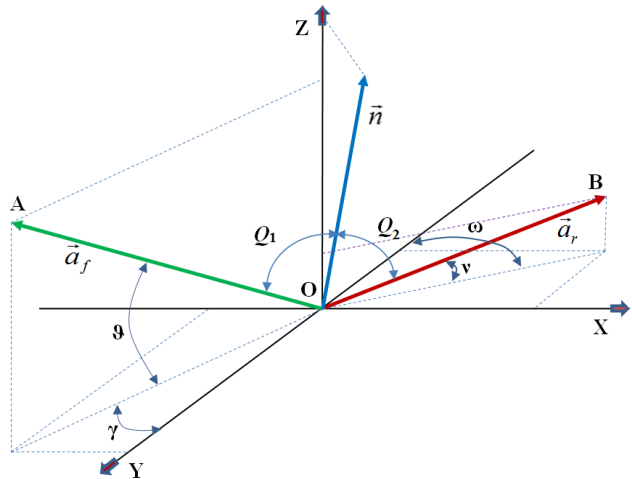

**Fig. A.1.2** Angles $\vartheta, \gamma$ determine the direction of the EMW ray incident on the facet;
angles $\nu, \omega$ determine the direction of the EMW ray reflected from the facet

$$E_\Sigma = \iint\limits_{b_n^2} E_i(x,y)dxdy = \frac{\sin\left\{\frac{2\pi b_n}{\lambda}(\cos\nu\sin\omega + \cos\vartheta\sin\gamma)\right\}}{\frac{2\pi b_n}{\lambda}(\cos\nu\sin\omega + \cos\vartheta\sin\gamma)} \times$$

$$\times \frac{\sin\left\{\frac{2\pi b_n}{\lambda}(\cos\nu\cos\omega + \cos\vartheta\cos\gamma)\right\}}{\frac{2\pi b_n}{\lambda}(\cos\nu\cos\omega + \cos\vartheta\cos\gamma)} \times \frac{E_m}{r_1}\exp\left\{i\left(\omega_1 t - \frac{2\pi}{\lambda}r_2\right)\right\}.$$

(A.1.2)

The first and second multipliers in the expression (A.1.2), squared, is the desired power scattering diagram of a flat, monochromatic EMW from a perfectly conducting square sections of the surface (facets)

$$D_r(\nu,\omega/\vartheta,\gamma) = \frac{\sin^2\left\{\frac{2\pi b_n}{\lambda}(\cos\nu\sin\omega + \cos\vartheta\sin\gamma)\right\}}{\left[\frac{2\pi b_n}{\lambda}(\cos\nu\sin\omega + \cos\vartheta\sin\gamma)\right]^2} \frac{\sin^2\left\{\frac{2\pi b_n}{\lambda}(\cos\nu\cos\omega + \cos\vartheta\cos\gamma)\right\}}{\left[\frac{2\pi b_n}{\lambda}(\cos\nu\cos\omega + \cos\vartheta\cos\gamma)\right]^2}.$$

(A.1.3)

The scattering diagrams calculated using the formula (A.1.3) are shown in Figure A.1.3 (see Appendix 12)

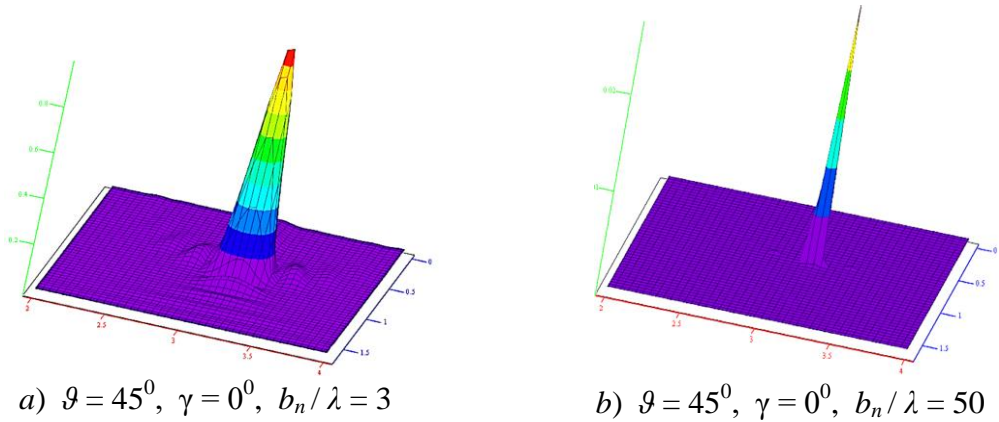

a) $\vartheta = 45^0$, $\gamma = 0^0$, $b_n/\lambda = 3$       b) $\vartheta = 45^0$, $\gamma = 0^0$, $b_n/\lambda = 50$

**Fig. A.1.3.** Power scattering diagrams of a flat, monochromatic EMW from a perfectly conducting square sections of the surface (facets). The calculations are performed according to the formula (A.1.3) using the MathCad software

The cross section of the scattering diagram (A.1.3) in the plane of incidence and reflection of the EMW beam shown in Figure A.1.4.

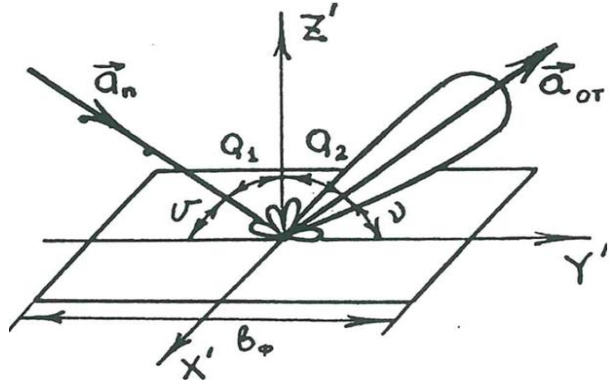

**Fig. A.1.4** Cross section of the scattering diagram of a beam flat electromagnetic wave from a flat square surface area (facet) conducting an electric current

From the scattering diagrams (DR) shown in Fig. A.1.3, it is seen that with an increase in the ratio $b_n / \lambda$, the main lobe of the DR becomes thinner and elongates, and the side lobes disappear. For large $b_n$ with respect to $\lambda$ (i.e., when $b_n / \lambda \to \infty$), the scattering diagram (A.1.3) degenerates into a delta function, i.e. the EMW beam reflected by the large facet becomes infinitely thin. In this case, the laws of reflection of a light ray from a facet (i.e., the laws of geometric optics) completely coincide with the laws of elastic reflection of particles from a solid surface under similar conditions (i.e., when the particles are much smaller than the dimensions of a solid surface).

In other words, in this case, the behavior of the light beam completely corresponds to the behavior of the particle (which can be called a photon). A photon is reflected almost lossless from a "mirror" surface according to the laws of geometric optics, just as elastic electrons or protons are reflected from a solid surface. Energy losses due to heating of the reflecting surface during collisions with particles and other secondary effects are not taken into account in the model under consideration.

Therefore, in this article, microparticles are any particles: fermions (e.g., electrons) and bosons (e.g., photons), whose sizes are much smaller than the characteristic irregularities of the reflecting surface (Kirchhoff approximation), and reflected from this surface according to the laws of geometric optics.

All conclusions made in this article relate to both elastic particles and electromagnetic radiation (light) rays, under the above conditions.

In connection with the foregoing, all conclusions made in this article relate to both elastic particles and EMW (light) rays, if the above conditions are met.

Appendix 2

## The Calculation of Integrals

Calculate the integrals (2.73)

$$\psi(\xi') = \frac{1}{\sqrt{2\pi}} \int_0^{l_2} \sqrt{\frac{2}{l_2}} \sin(\pi n_1 \xi / l_2) \exp\{i\xi'\xi/\eta\} d\xi, \quad (A.2.1)$$

$$\psi*(\xi') = \frac{1}{\sqrt{2\pi}} \int_0^{l_2} \sqrt{\frac{2}{l_2}} \sin(\pi n_1 \xi / l_2) \exp\{-i\xi'\xi/\eta\} d\xi \quad (A.2.2)$$

We start with the integral (A.2.1), and use the formula $\sin x = \dfrac{e^{ix} - e^{-ix}}{2i}$ and represent (A.2.1) in the form

$$\psi(\xi') = \frac{1}{\sqrt{2\pi}} \sqrt{\frac{2}{l_2}} \int_0^{l_2} \frac{e^{i\pi n_1 \xi / l_2} - e^{-i\pi n_1 \xi / l_2}}{2i} e^{i\xi'\xi/\eta} d\xi \quad (A.2.3)$$

Let's perform the following transformations

$$\psi(\xi') = \frac{1}{\sqrt{2\pi}} \sqrt{\frac{2}{l_2}} \int_0^{l_2} \frac{e^{i\pi n_1 \xi / l_2} e^{i\xi'\xi/\eta} - e^{-i\pi n_1 \xi / l_2} e^{i\xi'\xi/\eta}}{2i} d\xi$$

$$\psi(\xi') = \frac{1}{\sqrt{2\pi}} \sqrt{\frac{2}{l_2}} \int_{-\infty}^{\infty} \frac{e^{i\pi n_1 \xi / l_2 + i\xi'\xi/\eta} - e^{-i\pi n_1 \xi / l_2 + i\xi'\xi/\eta}}{2i} d\xi$$

$$\psi(\xi') = \frac{1}{\sqrt{2\pi}} \sqrt{\frac{2}{l_2}} \int_0^{l_2} \frac{e^{i\xi(\pi n_1 / l_2 + \xi'/\eta)} - e^{-i\xi(\pi n_1 / l_2 - \xi'/\eta)}}{2i} d\xi$$

$$\psi(\xi') = \frac{1}{2i\sqrt{2\pi}} \sqrt{\frac{2}{l_2}} \int_0^{l_2} e^{i\xi(\pi n_1 / l_2 + \xi'/\eta)} - e^{-i\xi(\pi n_1 / l_2 - \xi'/\eta)} d\xi$$

As a result of these transformations, we obtain

$$\psi(\xi') = \frac{1}{2i\sqrt{2\pi}} \sqrt{\frac{2}{l_2}} \left( \int_0^{l_2} e^{i\xi(\pi n_1/l_2 + \xi'/\eta)} d\xi - \int_0^{l_2} e^{-i\xi(\pi n_1/l_2 - \xi'/\eta)} d\xi \right) \quad (A.2.4)$$

Let's calculate the first integral in (A.2.4)

$$\int_{-\infty}^{\infty} e^{i\xi(\pi n_1\xi/l_2 + \xi'/\eta)} d\xi = \frac{\int_0^{l_2} e^{i\xi(\pi n_1/l_2 + \xi'/\eta)} d[i\xi(\pi n_1/l_2 + \xi'/\eta)]}{i(\pi n_1/l_2 + \xi'/\eta)}$$

$$\frac{\int_0^{l_2} e^{i\xi(\pi n_1\xi/l_2 + \xi'/\eta)} di\xi(\pi n_1/l_2 + \xi'/\eta)}{i(\pi n_1/l_2 + \xi'/\eta)} = \frac{e^{i\xi(\pi n_1/l_2 + \xi'/\eta)}}{i(\pi n_1/l_2 + \xi'/\eta)} \bigg|_0^{l_2}$$

$$\frac{e^{i\xi(\pi n_1/l_2 + \xi'/\eta)}}{i(\pi n_1/l_2 + \xi'/\eta)} \bigg|_0^{l_2} = \frac{e^{il_2(\pi n_1/l_2 + \xi'/\eta)}}{i(\pi n_1/l_2 + \xi'/\eta)} - \frac{e^{i0(\pi n_1/l_2 + \xi'/\eta)}}{i(\pi n_1/l_2 + \xi'/\eta)}$$

$$\frac{e^{i\xi(\pi n_1/l_2 + \xi'/\eta)}}{i(\pi n_1/l_2 + \xi'/\eta)} \bigg|_0^{l_2} = \frac{e^{il_2(\pi n_1/l_2 + \xi'/\eta)}}{i(\pi n_1/l_2 + \xi'/\eta)} - \frac{1}{i(\pi n_1/l_2 + \xi'/\eta)}$$

As a result of these calculations, we obtain

$$\frac{e^{i\xi(\pi n_1/l_2 + \xi'/\eta)}}{i(\pi n_1/l_2 + \xi'/\eta)} \bigg|_0^{l_2} = \frac{e^{il_2(\pi n_1/l_2 + \xi'/\eta)} - 1}{i(\pi n_1/l_2 + \xi'/\eta)} \quad (A.2.5)$$

Let's calculate the second integral in (A.2.4)

$$\int_0^{l_2} e^{-i\xi(\pi n_1\xi/l_2 - \xi'/\eta)} d\xi = \frac{\int_0^{l_2} e^{-i\xi(\pi n_1/l_2 - \xi'/\eta)} d[-i\xi(\pi n_1/l_2 - \xi'/\eta)]}{-i(\pi n_1/l_2 - \xi'/\eta)}$$

$$\frac{\int_0^{l_2} e^{-i\xi(\pi n_1/l_2 - \xi'/\eta)} d-i\xi(\pi n_1/l_2 - \xi'/\eta)}{-i(\pi n_1/l_2 - \xi'/\eta)} = \frac{e^{-i\xi(\pi n_1\xi/l_2 - \xi'/\eta)}}{-i(\pi n_1/l_2 - \xi'/\eta)} \bigg|_0^{l_2}$$

$$\frac{e^{-i\xi(\pi n_1/l_2 - \xi'/\eta)}}{-i(\pi n_1/l_2 - \xi'/\eta)} \bigg|_0^{l_2} = \frac{e^{-il_2(\pi n_1/l_2 - \xi'/\eta)}}{i(\pi n_1/l_2 - \xi'/\eta)} - \frac{1}{i(\pi n_1/l_2 - \xi'/\eta)}$$

As a result of these calculations, we obtain

$$\frac{e^{-i\xi(\pi n_1/l_2 - \xi'/\eta)}}{-i(\pi n_1/l_2 - \xi'/\eta)} \bigg|_0^{l_2} = \frac{e^{-il_2(\pi n_1/l_2 - \xi'/\eta)} - 1}{-i(\pi n_1/l_2 - \xi'/\eta)} \quad (A.2.6)$$

Substituting (A.2.5) and (A.2.6) and (A.2.4), we have

$$\psi(\xi') = \frac{1}{2i\sqrt{2\pi}}\sqrt{\frac{2}{l_2}}\left(\frac{e^{il_2(\pi n_1/l_2+\xi'/\eta)}-1}{i(\pi n_1/l_2+\xi'/\eta)} - \frac{e^{-il_2(\pi n_1/l_2-\xi'/\eta)}-1}{-i(\pi n_1/l_2-\xi'/\eta)}\right)$$

Let's do the transformations

$$\psi(\xi') = \frac{1}{2ii\sqrt{2\pi}}\sqrt{\frac{2}{l_2}}\left(\frac{e^{i(\pi n_1+\xi' l_2/\eta)}-1}{(\pi n_1/l_2+\xi'/\eta)} + \frac{e^{-i(\pi n_1-\xi' l_2/\eta)}-1}{(\pi n_1/l_2-\xi'/\eta)}\right)$$

$$\psi(\xi') = -\frac{1}{2\sqrt{2\pi}}\sqrt{\frac{2}{l_2}}\left(\frac{e^{i(\pi n_1+\xi' l_2/\eta)}-1}{(\pi n_1/l_2+\xi'/\eta)} + \frac{e^{-i(\pi n_1-\xi' l_2/\eta)}-1}{(\pi n_1/l_2-\xi'/\eta)}\right)$$

Finally we get the result of integration (A.2.1)

$$\psi(\xi') = -\sqrt{\frac{1}{4\pi l_2}}\left(\frac{e^{i(\pi n_1+\xi' l_2/\eta)}-1}{(\pi n_1/l_2+\xi'/\eta)} + \frac{e^{-i(\pi n_1-\xi' l_2/\eta)}-1}{(\pi n_1/l_2-\xi'/\eta)}\right) \quad (A.2.7)$$

Similarly, we take the integral (A.2.2)

$$\psi^*(\xi') = \frac{1}{\sqrt{2\pi}}\int_0^{l_2}\sqrt{\frac{2}{l_2}}\sin(\pi n_1\xi/l_2)\exp\{-i\xi'\xi/\eta\}d\xi$$

Let's represent (A.2.2) in the form

$$\psi^*(\xi') = \frac{1}{\sqrt{2\pi}}\sqrt{\frac{2}{l_2}}\int_0^{l_2}\frac{e^{i\pi n_1\xi/l_2}-e^{-i\pi n_1\xi/l_2}}{2i}e^{-i\xi'\xi/\eta}d\xi$$

$$\psi^*(\xi') = \frac{1}{\sqrt{2\pi}}\sqrt{\frac{2}{l_2}}\int_0^{l_2}\frac{e^{i\pi n_1\xi/l_2}e^{-i\xi'\xi/\eta}-e^{-i\pi n_1\xi/l_2}e^{-i\xi'\xi/\eta}}{2i}d\xi$$

$$\psi^*(\xi') = \frac{1}{\sqrt{2\pi}}\sqrt{\frac{2}{l_2}}\int_{-\infty}^{\infty}\frac{e^{i\pi n_1\xi/l_2-i\xi'\xi/\eta}-e^{-i\pi n_1\xi/l_2-i\xi'\xi/\eta}}{2i}d\xi$$

$$\psi^*(\xi') = \frac{1}{\sqrt{2\pi}}\sqrt{\frac{2}{l_2}}\int_0^{l_2}\frac{e^{i\xi(\pi n_1/l_2-\xi'/\eta)}-e^{-i\xi(\pi n_1/l_2+\xi'/\eta)}}{2i}d\xi$$

$$\psi^*(\xi') = \frac{1}{2i\sqrt{2\pi}}\sqrt{\frac{2}{l_2}}\int_0^{l_2}e^{i\xi(\pi n_1/l_2-\xi'/\eta)}-e^{-i\xi(\pi n_1/l_2+\xi'/\eta)}d\xi$$

As a result of these transformations, we obtain

$$\psi^*(\xi') = \frac{1}{2i\sqrt{2\pi}}\sqrt{\frac{2}{l_2}}\left(\int_0^{l_2}e^{i\xi(\pi n_1/l_2-\xi'/\eta)}d\xi - \int_0^{l_2}e^{-i\xi(\pi n_1/l_2+\xi'/\eta)}d\xi\right) \quad (A.2.8)$$

We calculate the first integral in (A.2.8)

$$\int_{-\infty}^{\infty} e^{i\xi(\pi n_1\xi/l_2+\xi'/\eta)}d\xi = \frac{\int_0^{l_2} e^{i\xi(\pi n_1/l_2-\xi'/\eta)}d[i\xi(\pi n_1/l_2-\xi'/\eta)]}{i(\pi n_1\xi/l_2-\xi'/\eta)}$$

$$\frac{\int_0^{l_2} e^{i\xi(\pi n_1\xi/l_2-\xi'/\eta)}di\xi(\pi n_1/l_2-\xi'/\eta)}{i(\pi n_1/l_2-\xi'/\eta)} = \frac{e^{i\xi(\pi n_1/l_2-\xi'/\eta)}}{i(\pi n_1/l_2-\xi'/\eta)}\Bigg|_0^{l_2}$$

$$\frac{e^{i\xi(\pi n_1/l_2-\xi'/\eta)}}{i(\pi n_1/l_2-\xi'/\eta)}\Bigg|_0^{l_2} = \frac{e^{il_2(\pi n_1/l_2-\xi'/\eta)}}{i(\pi n_1/l_2-\xi'/\eta)} - \frac{e^{i0(\pi n_1/l_2-\xi'/\eta)}}{i(\pi n_1/l_2-\xi'/\eta)}$$

$$\frac{e^{i\xi(\pi n_1/l_2-\xi'/\eta)}}{i(\pi n_1/l_2-\xi'/\eta)}\Bigg|_0^{l_2} = \frac{e^{il_2(\pi n_1/l_2-\xi'/\eta)}}{i(\pi n_1/l_2-\xi'/\eta)} - \frac{1}{i(\pi n_1/l_2-\xi'/\eta)}$$

$$\frac{e^{i\xi(\pi n_1/l_2-\xi'/\eta)}}{i(\pi n_1/l_2-\xi'/\eta)}\Bigg|_0^{l_2} = \frac{e^{il_2(\pi n_1/l_2-\xi'/\eta)}-1}{i(\pi n_1/l_2-\xi'/\eta)} \qquad (A.2.9)$$

We calculate the second integral in (A.2.8)

$$\int_0^{l_2} e^{-i\xi(\pi n_1\xi/l_2+\xi'/\eta)}d\xi = \frac{\int_0^{l_2} e^{-i\xi(\pi n_1/l_2+\xi'/\eta)}d[-i\xi(\pi n_1/l_2+\xi'/\eta)]}{-i(\pi n_1/l_2+\xi'/\eta)}$$

$$\frac{\int_0^{l_2} e^{-i\xi(\pi n_1/l_2+\xi'/\eta)}d-i\xi(\pi n_1/l_2+\xi'/\eta)}{-i(\pi n_1/l_2+\xi'/\eta)} = \frac{e^{-i\xi(\pi n_1\xi/l_2+\xi'/\eta)}}{-i(\pi n_1/l_2+\xi'/\eta)}\Bigg|_0^{l_2}$$

$$\frac{e^{-i\xi(\pi n_1/l_2+\xi'/\eta)}}{-i(\pi n_1/l_2+\xi'/\eta)}\Bigg|_0^{l_2} = \frac{e^{-il_2(\pi n_1/l_2+\xi'/\eta)}-1}{-i(\pi n_1/l_2+\xi'/\eta)} \qquad (A.2.10)$$

Substituting (A.2.5) and (A.2.6) and (A.2.4), we have

$$\psi^*(\xi') = \frac{1}{2i\sqrt{2\pi}}\sqrt{\frac{2}{l_2}}\left(\frac{e^{il_2(\pi n_1/l_2-\xi'/\eta)}-1}{i(\pi n_1/l_2-\xi'/\eta)} - \frac{e^{-il_2(\pi n_1/l_2+\xi'/\eta)}-1}{-i(\pi n_1/l_2+\xi'/\eta)}\right)$$

Let's do the transformations

$$\psi^*(\xi') = \frac{1}{2ii\sqrt{2\pi}}\sqrt{\frac{2}{l_2}}\left(\frac{e^{i(\pi n_1-\xi' l_2/\eta)}-1}{(\pi n_1/l_2-\xi'/\eta)} + \frac{e^{-i(\pi n_1+\xi' l_2/\eta)}-1}{(\pi n_1/l_2+\xi'/\eta)}\right)$$

$$\psi^*(\xi') = -\frac{1}{2\sqrt{2\pi}}\sqrt{\frac{2}{l_2}}\left(\frac{e^{i(\pi n_1-\xi'l_2/\eta)}-1}{(\pi n_1/l_2-\xi'/\eta)}+\frac{e^{-i(\pi n_1+\xi'l_2/\eta)}-1}{(\pi n_1/l_2+\xi'/\eta)}\right)$$

Finally, we obtain the result of integration (A.2.2)

$$\psi^*(\xi') = -\sqrt{\frac{1}{4\pi l_2}}\left(\frac{e^{i(\pi n_1-\xi'l_2/\eta)}-1}{(\pi n_1/l_2-\xi'/\eta)}+\frac{e^{-i(\pi n_1+\xi'l_2/\eta)}-1}{(\pi n_1/l_2+\xi'/\eta)}\right) \qquad (A.2.11)$$

So, the results of taking the integrals (A.2.1) and (A.2.2) are the expressions (A.2.7) and (A.2.11):

$$\psi(\xi') = -\sqrt{\frac{1}{4\pi l_2}}\left(\frac{e^{i(\pi n_1+\xi'l_2/\eta)}-1}{(\pi n_1/l_2+\xi'/\eta)}+\frac{e^{-i(\pi n_1-\xi'l_2/\eta)}-1}{(\pi n_1/l_2-\xi'/\eta)}\right) \qquad (A.2.12)$$

$$\psi^*(\xi') = -\sqrt{\frac{1}{4\pi l_2}}\left(\frac{e^{i(\pi n_1-\xi'l_2/\eta)}-1}{(\pi n_1/l_2-\xi'/\eta)}+\frac{e^{-i(\pi n_1+\xi'l_2/\eta)}-1}{(\pi n_1/l_2+\xi'/\eta)}\right) \qquad (A.2.13)$$

Appendix 3

**The product of factors**

Product of expressions (2.74) and (2.75) {or (A.2.12) and (A.2.13)}

$$\psi(\xi') = -\sqrt{\frac{1}{4\pi l_2}}\left(\frac{e^{i(\pi n_1+\xi'l_2/\eta)}-1}{(\pi n_1/l_2+\xi'/\eta)}+\frac{e^{-i(\pi n_1-\xi'l_2/\eta)}-1}{(\pi n_1/l_2-\xi'/\eta)}\right) \qquad (A.3.1)$$

$$\psi^*(\xi') = -\sqrt{\frac{1}{4\pi l_2}}\left(\frac{e^{i(\pi n_1-\xi'l_2/\eta)}-1}{(\pi n_1/l_2-\xi'/\eta)}+\frac{e^{-i(\pi n_1+\xi'l_2/\eta)}-1}{(\pi n_1/l_2+\xi'/\eta)}\right) \qquad (A.3.2)$$

equally

$$p(\xi') = \psi(\xi')\psi^*(\xi') = \frac{1}{4\pi l_2}\left(\frac{e^{i(\pi n_1+\xi'l_2/\eta)}-1}{(\pi n_1/l_2+\xi'/\eta)}+\frac{e^{-i(\pi n_1-\xi'l_2/\eta)}-1}{(\pi n_1/l_2-\xi'/\eta)}\right)\left(\frac{e^{i(\pi n_1-\xi'l_2/\eta)}-1}{(\pi n_1/l_2-\xi'/\eta)}+\frac{e^{-i(\pi n_1+\xi'l_2/\eta)}-1}{(\pi n_1/l_2+\xi'/\eta)}\right)$$

$$(A.3.3)$$

Opening large brackets, we multiply the terms in pairs

$$\frac{e^{i(\pi n_1 + \xi' l_2/\eta)} - 1}{(\pi n_1/l_2 + \xi'/\eta)} \frac{e^{i(\pi n_1 - \xi' l_2/\eta)} - 1}{(\pi n_1/l_2 - \xi'/\eta)} = \frac{\left(e^{i(\pi n_1 + \xi' l_2/\eta)} - 1\right)\left(e^{i(\pi n_1 - \xi' l_2/\eta)} - 1\right)}{\left(\pi n_1/l_2\right)^2 - \left(\xi'/\eta\right)^2}$$

$$\frac{e^{i(\pi n_1 + \xi' l_2/\eta)} - 1}{(\pi n_1/l_2 + \xi'/\eta)} \frac{e^{-i(\pi n_1 + \xi' l_2/\eta)} - 1}{(\pi n_1/l_2 + \xi'/\eta)} = \frac{\left(e^{i(\pi n_1 + \xi' l_2/\eta)} - 1\right)\left(e^{-i(\pi n_1 + \xi' l_2/\eta)} - 1\right)}{\left(\pi n_1/l_2 + \xi'/\eta\right)^2}$$

$$\frac{e^{i(\pi n_1 + \xi' l_2/\eta)} - 1}{(\pi n_1/l_2 + \xi'/\eta)} \frac{e^{-i(\pi n_1 + \xi' l_2/\eta)} - 1}{(\pi n_1/l_2 + \xi'/\eta)} = \frac{\left(e^{i(\pi n_1 + \xi' l_2/\eta)} - 1\right)\left(e^{-i(\pi n_1 + \xi' l_2/\eta)} - 1\right)}{\left(\pi n_1/l_2 + \xi'/\eta\right)^2}$$

$$\frac{e^{-i(\pi n_1 - \xi' l_2/\eta)} - 1}{(\pi n_1/l_2 - \xi'/\eta)} \frac{e^{-i(\pi n_1 + \xi' l_2/\eta)} - 1}{(\pi n_1/l_2 + \xi'/\eta)} = \frac{\left(e^{-i(\pi n_1 - \xi' l_2/\eta)} - 1\right)\left(e^{-i(\pi n_1 + \xi' l_2/\eta)} - 1\right)}{\left(\pi n_1/l_2\right)^2 - \left(\xi'/\eta\right)^2}$$

Add the resulting expressions

$$\frac{\left(e^{i(\pi n_1 + \xi' l_2/\eta)} - 1\right)\left(e^{i(\pi n_1 - \xi' l_2/\eta)} - 1\right)}{\left(\pi n_1/l_2\right)^2 - \left(\xi'/\eta\right)^2} + \frac{2\left(e^{i(\pi n_1 + \xi' l_2/\eta)} - 1\right)\left(e^{-i(\pi n_1 + \xi' l_2/\eta)} - 1\right)}{\left(\pi n_1/l_2 + \xi'/\eta\right)^2} + \frac{\left(e^{-i(\pi n_1 - \xi' l_2/\eta)} - 1\right)\left(e^{-i(\pi n_1 + \xi' l_2/\eta)} - 1\right)}{\left(\pi n_1/l_2\right)^2 - \left(\xi'/\eta\right)^2}$$

Rearranging terms and summing up them, we get

$$\frac{\left(e^{i(\pi n_1 + \xi' l_2/\eta)} - 1\right)\left(e^{i(\pi n_1 - \xi' l_2/\eta)} - 1\right) + \left(e^{-i(\pi n_1 - \xi' l_2/\eta)} - 1\right)\left(e^{-i(\pi n_1 + \xi' l_2/\eta)} - 1\right)}{\left(\pi n_1/l_2\right)^2 - \left(\xi'/\eta\right)^2} + \frac{2\left(e^{i(\pi n_1 + \xi' l_2/\eta)} - 1\right)\left(e^{-i(\pi n_1 + \xi' l_2/\eta)} - 1\right)}{\left(\pi n_1/l_2 + \xi'/\eta\right)^2}$$

(A.3.4)

Performing calculations

1. $\left(e^{i(\pi n_1+\xi' l_2/\eta)}-1\right)\left(e^{i(\pi n_1-\xi' l_2/\eta)}-1\right)=e^{i2\pi n_1}-e^{i(\pi n_1+\xi' l_2/\eta)}-e^{i(\pi n_1-\xi' l_2/\eta)}+1$

2. $\left(e^{-i(\pi n_1-\xi' l_2/\eta)}-1\right)\left(e^{-i(\pi n_1+\xi' l_2/\eta)}-1\right)=e^{-i2\pi n_1}-e^{-i(\pi n_1-\xi' l_2/\eta)}-e^{-i(\pi n_1+\xi' l_2/\eta)}+1$

3. $\left(e^{i(\pi n_1+\xi' l_2/\eta)}-1\right)\left(e^{-i(\pi n_1+\xi' l_2/\eta)}-1\right)=e^0-e^{i(\pi n_1+\xi' l_2/\eta)}-e^{-i(\pi n_1+\xi' l_2/\eta)}+1=$

$=1-(e^{i(\pi n_1+\xi' l_2/\eta)}+e^{-i(\pi n_1+\xi' l_2/\eta)})+1=-(e^{i(\pi n_1+\xi' l_2/\eta)}+e^{-i(\pi n_1+\xi' l_2/\eta)})+2=$

$-2[(e^{i(\pi n_1+\xi' l_2/\eta)}+e^{-i(\pi n_1+\xi' l_2/\eta)})/2-1]=-2[\cos(\pi n_1+\xi' l_2/\eta)-1]$ \hfill (A.3.5)

where the expression $\cos x=\dfrac{e^{ix}+e^{-ix}}{2}$ is taken into account.

Add 1 and 2

$$e^{i2\pi n_1}-e^{i(\pi n_1+\xi' l_2/\eta)}-e^{i(\pi n_1-\xi' l_2/\eta)}+e^{-i2\pi n_1}-e^{-i(\pi n_1-\xi' l_2/\eta)}-e^{-i(\pi n_1+\xi' l_2/\eta)}+2$$

Let's regroup the terms

$$(e^{i2\pi n_1}+e^{-i2\pi n_1})-(e^{i(\pi n_1+\xi' l_2/\eta)}+e^{-i(\pi n_1+\xi' l_2/\eta)})-(e^{i(\pi n_1-\xi' l_2/\eta)}+e^{-i(\pi n_1-\xi' l_2/\eta)})+2$$

or

$$2[(e^{i2\pi n_1}+e^{-i2\pi n_1})/2-(e^{i(\pi n_1+\xi' l_2/\eta)}+e^{-i(\pi n_1+\xi' l_2/\eta)})/2-(e^{i(\pi n_1-\xi' l_2/\eta)}+e^{-i(\pi n_1-\xi' l_2/\eta)})/2+1]=$$

$=2[\cos 2\pi n_1-\cos(\pi n_1+\xi' l_2/\eta)-\cos(\pi n_1-\xi' l_2/\eta)+1]$ \hfill (A.3.6)

Let's substitute the terms (A.3.5) and (A.3.6) into (A.3.4), we obtain

$$\frac{2[(\cos 2\pi n_1-\cos(\pi n_1+\xi' l_2/\eta)-\cos(\pi n_1-\xi' l_2/\eta)+1]}{\left(\pi n_1/l_2\right)^2-\left(\xi'/\eta\right)^2}-\frac{4[\cos(\pi n_1+\xi' l_2/\eta)-1]}{\left(\pi n_1/l_2+\xi'/\eta\right)^2} \qquad (A.3.7)$$

Now insert (A.3.7) into (A.3.3)

$$p(\xi') = \frac{1}{4\pi l_2}\left(\frac{2[(\cos 2\pi n_1 - \cos(\pi n_1 + \xi' l_2/\eta) - \cos(\pi n_1 - \xi' l_2/\eta) + 1]}{\left(\pi n_1/l_2\right)^2 - \left(\xi'/\eta\right)^2} - \frac{4[\cos(\pi n_1 + \xi' l_2/\eta) - 1]}{\left(\pi n_1/l_2 + \xi'/\eta\right)^2}\right)$$

(A.3.8)

We use two trigonometric formulas

$$\cos^2 x = \frac{1+\cos 2x}{2} \quad \text{и} \quad \cos x \cos y = \frac{1}{2}[\cos(x-y) + \cos(x+y)] \quad (A.3.9)$$

Where should

$$\cos 2\pi n_1 + 1 = 2\cos^2 \pi n_1 \quad (A.3.10)$$

$$\cos(\pi n_1 - \xi' l_2/\eta) + \cos(\pi n_1 + \xi' l_2/\eta) = 2\cos(\pi n_1)\cos(\xi' l_2/\eta) \quad (A.3.11)$$

In view of (A.3.10) and (A.3.11), the expression (A.3.8) takes the form

$$p(\xi') = \frac{1}{4\pi l_2}\left(\frac{2[2\cos^2 \pi n_1 - 2\cos(\pi n_1)\cos(\xi' l_2/\eta)]}{\left(\pi n_1/l_2\right)^2 - \left(\xi'/\eta\right)^2} - \frac{4[\cos(\pi n_1 + \xi' l_2/\eta) - 1]}{\left(\pi n_1/l_2 + \xi'/\eta\right)^2}\right)$$

Performing simplifications

$$p(\xi') = \frac{1}{4\pi l_2}\left(\frac{4[\cos^2 \pi n_1 - \cos(\pi n_1)\cos(\xi' l_2/\eta)]}{\left(\pi n_1/l_2\right)^2 - \left(\xi'/\eta\right)^2} - \frac{4[\cos(\pi n_1 + \xi' l_2/\eta) - 1]}{\left(\pi n_1/l_2 + \xi'/\eta\right)^2}\right)$$

finally get

$$p(\xi') = \frac{1}{\pi l_2}\left(\frac{[\cos^2 \pi n_1 - \cos(\pi n_1)\cos(\xi' l_2/\eta)]}{\left(\pi n_1/l_2\right)^2 - \left(\xi'/\eta\right)^2} - \frac{[\cos(\pi n_1 + \xi' l_2/\eta) - 1]}{\left(\pi n_1/l_2 + \xi'/\eta\right)^2}\right) \quad (A.3.12)$$

## 7 Abbreviations and Definitions

DESM is diagram of elastic scattering of microparticles;

EDP is electron diffraction pattern;

OPDF is one-dimensional probability density function;

SD is standard deviation;

SRP is stationary random process;

TPDF is two-dimensional (or joint) probability density function.

*Microparticle* is a solid elastic compact body or a ray of light (i.e., a photon) whose size or wavelength is much smaller than the characteristic size of the irregularities of the reflecting surface, upon collision with which they are reflected according to the laws of geometric optics (see §1).

*Elastic scattering* is the reflection of a particle from a surface according to the laws of geometric optics (see Figures 3, 4): 1) The reflection of an elastic particle from a solid surface occurs in the plane of its incidence; 2) The angle of reflection $Q_2$ is equal to the angle of incidence $Q_1$.

## 8 Competing Interestsf

The author states that there are no competing interests.

## 9 Consent to publication

There is permission from the Moscow Aviation Institute for publication.

## 10 Ethics approval and consent to participate

Not applicable.

## 11 Funding

There is no funding for this work.

## 12 Data availability

Data confirming the results of this study can be obtained from the author of this article upon request at: alsignat@yandex.ru and at http://metraphysics.ru/.